\documentclass[lettersize,journal]{IEEEtran}
\usepackage{amsmath,amsfonts}
\usepackage{algorithmic}
\usepackage{algorithm}
\usepackage{array}
\usepackage[caption=false,font=normalsize,labelfont=sf,textfont=sf]{subfig}
\usepackage{textcomp}
\usepackage{stfloats}
\usepackage{url}
\usepackage{verbatim}
\usepackage{graphicx}
\usepackage{cite}
\usepackage{multirow}
\usepackage{bbding}
\usepackage{utfsym}
\usepackage{ulem}
\usepackage{balance}
\begin{document}

\title{Diffsound: Discrete Diffusion Model for Text-to-sound Generation}
%~\IEEEmembership{student member,~IEEE,}
\author{Dongchao Yang, Jianwei Yu, Helin Wang, Wen Wang, Chao Weng, Yuexian Zou,~\IEEEmembership{Senior Member,~IEEE} and Dong Yu,~\IEEEmembership{Fellow,~IEEE}
        % <-this % stops a space
\thanks{Dongchao Yang, Helin Wang, Wen Wang and Yuexian Zou are with the Advanced Data and Signal Processing laboratory, School of Electronic and Computer Engineering, Peking University, China. This work was done when Dongchao Yang was an intern at Tencent AI Lab.}% <-this % stops a space
\thanks{Jianwei Yu, Chao Weng and Dong Yu are with Tencent AI Lab.}
\thanks{Yuexian Zou and Jianwei Yu are the corresponding authors.\\ 
(zouyx@pku.edu.cn; tomasyu@tencent.com)}}

\markboth{Journal of \LaTeX\ Class Files,~Vol.~14, No.~8, August~2021}%
{Shell \MakeLowercase{\textit{et al.}}: A Sample Article Using IEEEtran.cls for IEEE Journals}


\maketitle
% Generating sound effects that humans want is an important topic. Many works have been proposed to generate sound conditioned on a one-hot label or a video. However, it is a labor-extensive and time-consuming process to generate sound according to video information, because video data is huge and difficult to obtain and access. Using the one-hot label as conditional information is an efficient way, but it is hard to extend to the open domain. In contrast, text information is content-rich and easy to access. which leads to generating low-quality sound
% There are two challenges that exist in text-to-sound generation tasks. Firstly, most of the sound generation methods employ the autoregressive (AR) decoder to generate the mel-spectrogram based on the conditional information obtained from a encoder, and then a vocoder is used to transform the generated mel-spectrogram into waveform. However,
% Secondly, there is only a small-scale text-audio dataset (\textit{e.g.} Audiocaps).
% To solve the data deficiency problem in the text-to-sound generation task, we propose a mask-based text generation strategy (MBTG) to utilize the large-scale audio events dataset (\textbf{e.g.} Audioset), which only includes the category label of the audio. We demonstrate that the Diffsound pre-trained on Audioset can be easily finetuned on a small-scale text-audio dataset (\textit{e.g.} Audiocaps).
% \sout{has been proved as a state-of-the-art method}
\begin{abstract} Generating sound effects that people want is an important topic. However, there are limited studies in this area for sound generation. In this study, we investigate generating sound conditioned on a text prompt and propose a novel text-to-sound generation framework that consists of a text encoder, a Vector Quantized Variational Autoencoder (VQ-VAE), a token-decoder, and a vocoder. The framework first uses the token-decoder to transfer the text features extracted from the text encoder to a mel-spectrogram with the help of VQ-VAE, and then the vocoder is used to transform the generated mel-spectrogram into a waveform. We found that the token-decoder significantly influences the generation performance. Thus, we focus on designing a good token-decoder in this study. We begin with the traditional {\color{black}{autoregressive (AR)}} token-decoder, which has shown state-of-the-art performance in previous sound generation works. However, the AR token-decoder always predicts the mel-spectrogram tokens one by one in order, which may introduce the unidirectional bias and accumulation of errors problems. Moreover, with the AR token-decoder, the sound generation time increases linearly with the sound duration. To overcome the shortcomings introduced by AR token-decoders, we propose a non-autoregressive token-decoder based on the discrete diffusion model, named Diffsound. Specifically, the Diffsound model predicts all of the mel-spectrogram tokens in one step and then refines the predicted tokens in the next step, so the best-predicted results can be obtained by iteration. Our experiments show that our proposed Diffsound model not only produces better text-to-sound generation results when compared with the AR token-decoder but also has a faster generation speed, \textit{i.e.}, MOS: 3.56 \textit{v.s} 2.786, and the generation speed is five times faster than the AR decoder. Furthermore, to automatically assess the quality of generated samples, we define three different objective evaluation metrics \textit{i.e.}, Fréchet Inception Distance (FID), Kullback-Leibler (KL), and audio caption loss, which can comprehensively assess the relevance and fidelity of the generated samples. 
Code, pre-trained models, and generated samples are released \footnote{http://dongchaoyang.top/text-to-sound-synthesis-demo/}.

\end{abstract}

% Furthermore, one of the major challenges in audio analysis fields is data deficient. The generated data can be viewed as extra training data for sound event classification and detection tasks.
% Many models have been proposed to controllably generate \textit{e.g.} images \cite{ding2021cogview,dhariwal2021diffusion,esser2021taming,reed2016generative,chang2022maskgit,nichol2021glide}, videos \cite{chan2019everybody,hao2018controllable,mallya2020world,tulyakov2018mocogan}, and audios \cite{bazin2021spectrogram,chen2017deep,dhariwal2020jukebox,hao2018cmcgan,nistal2020drumgan,tomczak2020drum}, or separate sounds \cite{liu2021conditional,chen2020generating,zhou2018visual,iashin2021taming}. However, most of the audio works are music-related,
\section{Introduction}
\begin{figure}[t]
  \centering
  \includegraphics[width=\linewidth]{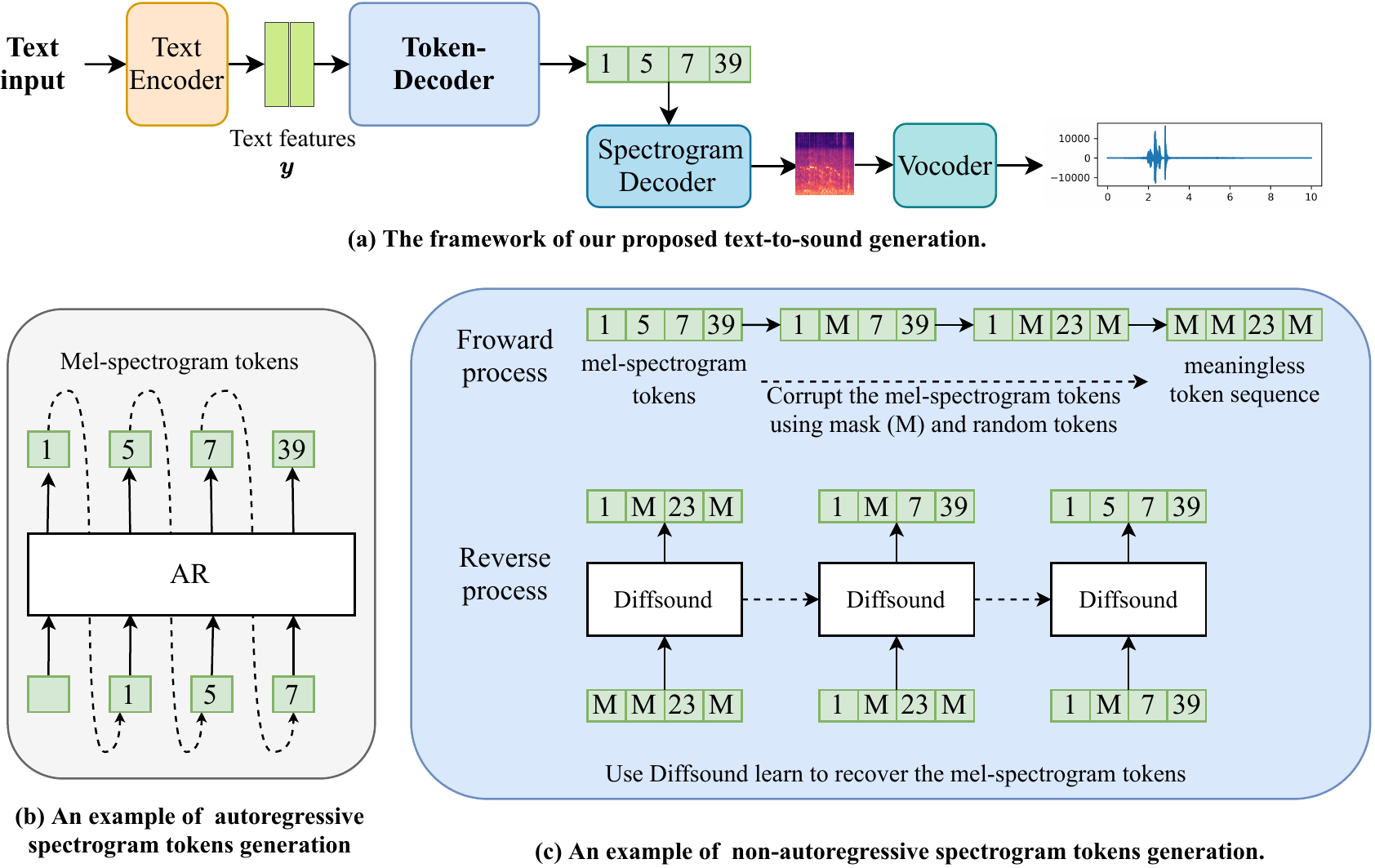}
  \caption{(a) shows the diagram of the text-to-sound generation framework includes four parts: a text encoder that extracts text features from the text input, a token-decoder that generates mel-spectrogram tokens, a pre-trained spectrorgam decoder that transforms the tokens into mel-spectrogram, and a vocoder that transforms the generated mel-spectrogram into waveform. We explore two kinds of token-decoders, an autoregressive (AR) token-decoder and a non-autoregressive token-decoder (Diffsound). (b) and (c) show the examples of AR token-decoder and non-AR token-decoder.}
  \label{fig:1}
  \vspace*{-\baselineskip}
\end{figure}

\IEEEPARstart{U}{ser} controlled sound generation has a lot of potential applications, such as movie and music productions, game scene sound effects, and so on. With the development of virtual reality (VR) technology, it is very important to generate the sound effects that users want.
% For example, many movie designers are required to search through large databases of sound effects to find a suitable sound for a scene. A less labor-extensive approach would be to automatically generate a novel and relevant sound, given a few text cues. 
% However, generating sound conditions on one-hot label is hard to extend to open domain sound generation.
% generating sound conditions on video is very time-consuming and labor-extensive due to video data being huge and hard to access, (\textit{e.g.} a 10-second video requires approximately 5Mb of storage). Using a one-hot label as conditional information is an efficient way, but it is hard to extend to open domain sound generation.
% which is content-rich and easy to obtain, and we can describe any sound in a single sentence.
Research on sound generation is very limited. Chen \textit{et al.} \cite{chen2020generating}, Zhou \textit{et al.} \cite{zhou2018visual} and Iashin and Rahtu \cite{iashin2021taming} proposed to generate sound related to a video. Liu \textit{et al.} \cite{liu2021conditional} and Kong \textit{et al.} \cite{kong2019acoustic} attempted to generate environmental sound conditioned on a one-hot label. 
{\color{black}{However, at the time of this work, there are very limited published works on generating sound from text descriptions. To the best of our knowledge, this paper is among the first work in this direction. }}
% \sout{However, at the time of this work, no works had been published on generating sound from text descriptions.}
Text-to-sound generation has a wide range of applications, \textit{e.g.}, adding background sound for speech synthesis systems. Nowadays, speech synthesis systems have been applied to poetry or novel reading. The user experience could be improved by adding background sound to scenarios represented in text. Furthermore, many music or movie designers are required to find a suitable sound for a scene. A simple approach is that they describe the scene with a sentence, and then use the text-to-sound model to generate the corresponding sound. In this work, we focus on directly generating audio based on human-written descriptions, such as ``An audience cheers and applauds while a man talks''.
%or ``Several birds tweet and ducks quack far away''. 
% \sout{a sequence containing a fixed number of tokens}
The state-of-the-art methods \cite{iashin2021taming,liu2021conditional} in the sound generation both employ a two-stage generation strategy, which first uses autoregressive (AR) decoder to generate a mel-spectrogram conditioned on a one-hot label or a video and then employs a vocoder (\textit{e.g.} MelGAN \cite{kumar2019melgan}) to transform the generated mel-spectrogram into waveform. To improve the generation efficiency, they propose to learn a prior in the form of the Vector Quantized Variational Autoencoder (VQ-VAE) codebook \cite{van2017neural}, which aims to compress the mel-spectrogram into  {\color{black}{a token sequence}}. 
% \sout{By using VQ-VAE, the mel-spectrogram generation problem transfers to predicting a sequence of discrete tokens corresponding to the mel-spectrogram.} 
{\color{black} With VQ-VAE, the mel-spectrogram generation problem can be formulated as predicting a sequence of discrete tokens from the text inputs.}
Inspired by \cite{iashin2021taming,liu2021conditional}, we propose a text-to-sound generation framework, which consists of a text encoder, a VQ-VAE, a token-decoder, and a vocoder. The diagram of the text-to-sound framework is shown in Figure \ref{fig:1}. We found that the token-decoder significantly influences the generation performance. Thus, we focus on designing a good token-decoder in this paper.
We start by looking into the AR token-decoder. However, we discovered that the AR token-decoder is unable to produce high-fidelity and high-relevance sound with text input. {\color{black}Though the AR token-decoder has been widely adopted} in sound generation tasks in previous research \cite{liu2021conditional,iashin2021taming}, it has two flaws: (1) Mel-spectrogram tokens are always predicted in order (\textit{e.g.}, from left to right) by the AR token-decoder. 
{\color{black} Such unidirectional predictions may restrict the sound generation performance to be sub-optimal since the information of a specific sound event location may come from both the left and the right context;}
% \sout{The fixed order results in model restricts the expressivity for time-frequency spectrogram modeling since the prediction of a specific location should not merely attend to the context on the left, but also attend to right;} 
(2) During the inference phase, incorrectly predicted tokens from previous steps propagate to subsequent tokens, resulting in accumulated prediction errors. Another issue in the text-to-sound generation is lacking text-audio pairs. The largest public available text-audio dataset is AudioCaps\cite{kim2019audiocaps}, which only includes about 49K text-audio samples. In contrast, Iashin and Rahtu \cite{iashin2021taming} trains their model using VGGSound dataset \cite{chen2020vggsound}, which has over 200K audio-video pairs.
% the AR models will lead to the accumulated prediction errors. Each step of the inference stage is performed based on previously sampled tokens – this is different from that of the training stage, which relies on the so-called ``teacher-forcing'' practice \cite{esser2021imagebart} and provides the ground truth for each step. This difference is important and its consequence merits careful examination. In particular, a token in the inference stage, once predicted, cannot be corrected and its errors will propagate to the subsequent tokens. 

To address the weaknesses of the AR token-decoder, we propose a non-autoregressive token-decoder based on diffusion probabilistic models (diffusion models for short) \cite{ho2020denoising,sohl2015deep,austin2021structured,gu2021vector}, named Diffsound. Instead of predicting the mel-spectrogram tokens one by one in order, Diffsound model predicts all of the mel-spectrogram tokens simultaneously, then it revises the previous predicted results in the following steps, so that the best results can be obtained by iterations. In each step, the Diffsound model leverages the contextual information of all tokens predicted in the previous step to estimate a new probability density distribution and uses this distribution to sample the tokens in the current step. Due to the fact that Diffsound model can make use of the contextual information of all tokens and revise any token in each step, we speculate that it can effectively alleviate  the unidirectional bias and the accumulated prediction error problems. We adopt the idea from diffusion models, which use a forward process to corrupt the original mel-spectrogram tokens in $T$ steps, and then let the model learn to recover the original tokens in a reverse process. Specifically, in the forward process, we define a transition matrix that denotes probability of each token transfer to a random token or a pre-defined MASK token. By using the transition matrix, the original tokens $\boldsymbol{x}_0 \sim q(\boldsymbol{x}_0)$ transfer into a stationary distribution $p(\boldsymbol{x}_T)$. In the reverse process, we let the network learn to recover the original tokens from $\boldsymbol{x}_T \sim p(\boldsymbol{x}_T)$ conditioned on the text features. Figure \ref{fig:1} (c) shows an example of non-autoregressive mel-spectrogram tokens generation.
%Furthermore, we can see that if the timesteps is smaller than the number of the spectrogram tokens, Diffsound will have a faster generation speed than AR model.

% \sout{The process of collecting text-audio pair is very expensive. Thus, data deficiency is one of the problem in the text-to-sound generation tasks.}

{\color{black}To address the problem of lacking text-audio pairs,} we propose to let the Diffsound model learn knowledge from the AudioSet dataset \cite{gemmeke2017audio} and then fine-tune the pre-trained Diffsound model on a small-scale text-audio dataset (\textit{e.g.}, AudioCaps). AudioSet is the largest available dataset in the audio field, but it only provides the event labels for each audio clip. To utilize the AudioSet dataset, we propose a mask-based text generation strategy (MBTG) that can generate a text description according to the event labels so that a new text-audio dataset is built. Furthermore, we observe a phenomenon: it is easier to generate audio that only includes one single event than audio that includes multiple events. To help the Diffsound model learn better, we mimic the human learning process by letting the Diffsound model learn from easy {\color{black}clips} and gradually advance to complex clips and knowledge. Specifically, we propose a curriculum learning strategy in our pre-training stage, that is, we first select the {\color{black}audio clips} that only include one event (easy sample) to the training set, and gradually add the audio clips that include multiple events (hard sample) to the training set.  
% This bidirectional attention provides global context for each token prediction and eliminates the unidirectional bias existed in AR model. 
% In the inference stage, we update the density distribution of all tokens in each step and resample all tokens according to the new distribution. Thus we can modify the wrong tokens and prevent error accumulation.

Human evaluation of sound generation models is an expensive and tedious procedure. Thus, objective evaluation metrics are necessary for sound generation tasks. We explore three objective evaluation metrics: Frechet Inception Distance (FID) \cite{heusel2017gans}, KL-divergence \cite{iashin2021taming} and audio caption loss. We demonstrate that these metrics can effectively evaluate the fidelity and relevance of the generated sound. Furthermore, we also use the Mean Opinion Score (MOS) to assess our methods. 

Experiments show that our text-to-sound generation framework can generate high-quality sound, \textit{e.g.}, MOS: 3.56 (ours) \textit{v.s} 4.11 (ground truth), and our proposed Diffsound model has better generation performance and speed compared to the AR token-decoder, \textit{e.g.}, MOS: 3.56 \textit{v.s} 2.786, and the generation speed is five times faster than the AR token-decoder. Our main contributions are listed as follows:
\begin{figure}[t]
  \centering
  \includegraphics[width=0.9\linewidth]{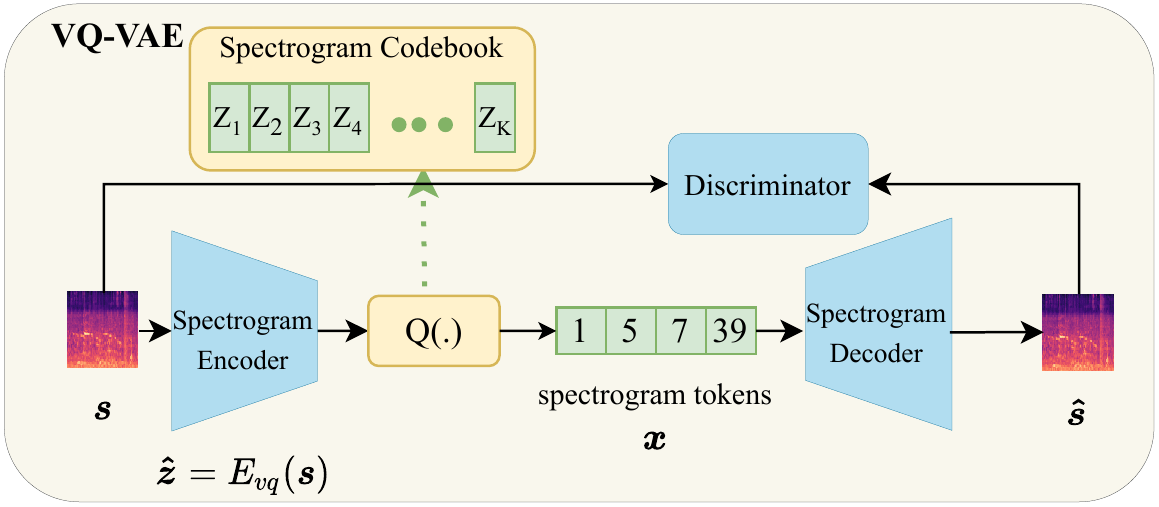}
  \caption{The overall architecture of VQ-VAE, which consists of four parts: an spectrogram encoder that extracts the representation $\boldsymbol{\hat{z}}$ from the mel-spectrogram, a codebook that contains a finite number of embedding vectors, a spectrogram decoder that reconstructs the mel-spectrogram based on mel-spectrogram tokens, and a discriminator that distinguishes the mel-spectrogram is original or reconstructed. $Q(.)$ denotes a quantizer that maps each features $\boldsymbol{\hat{z}}_{ij}$ into its closest codebook entry $\boldsymbol{z}_k$ to obtain the mel-spectrogram tokens.}
  \label{fig:gan}
  \vspace*{-\baselineskip}
\end{figure}

(1) For the first time, we investigate how to generate sound based on text descriptions and offer a text-to-sound generation framework. Furthermore, we propose a novel token-decoder (Diffsound) based on a discrete diffusion model that outperforms the AR token-decoder in terms of generation performance and speed.

(2) To solve the problem of lacking text-audio pairs in the text-to-sound generation task, we propose a mask-based text generation strategy (MBTG), which helps build a large-scale text-audio dataset based on the AudioSet dataset. We demonstrate the effectiveness of pre-training the Diffsound on the AudioSet, which gives the insight to improve the performance of the generation model under a data deficiency situation.

(3) We initially explore three objective evaluation metrics for the text-to-sound generation task. 
{\color{black}We show that these metrics can comprehensively evaluate the generated sound's relevance (\textit{i.e.}, how well the audio content relates to the text descriptions), as well as its fidelity (\textit{i.e.}, how closely the generated audio clip resembles actual environmental sound).}
\section{Related work}
 \noindent \textbf{GAN-based Content Generation} In the past few years, Generative Adversarial Networks (GANs) have shown promising results in image generation \cite{reed2016generative,zhang2018photographic,brock2018large}, speech synthesis \cite{kumar2019melgan,lee2020multi,guo2019new} and music generation \cite{nistal2020drumgan}. GAN-based models are capable of synthesizing high-fidelity images/sounds. However, they suffer from well-known training instability and mode collapse issues, which lead to a lack of sample diversity. Most related to our work is RegNet \cite{chen2020generating}, which aims to generate sound conditioned by visual information. It is noted that RegNet only generates sounds on a single domain dataset (\textit{e.g.}, dog bark or drum), which means that it struggles on complex scenes with multiple sound events. Furthermore, Iashin and Rahtu \cite{iashin2021taming} have demonstrated that using an autoregressive generation model can produce better results than RegNet.
 
 \noindent  \textbf{Autoregressive Models} AR models \cite{brown2020language,radford2018improving} have shown powerful generation capability and have been applied for image generation \cite{esser2021taming,ding2021cogview,van2017neural,parmar2018image,razavi2019generating,ramesh2021zero,AdityaRamesh2022HierarchicalTI}, speech synthesis \cite{wang2017tacotron} and sound generation \cite{liu2021conditional,iashin2021taming}. To generate high-resolution images, VQ-VAE \cite{van2017neural,razavi2019generating}, VQGAN \cite{esser2021taming} and ImageBART \cite{esser2021imagebart}
train an encoder to compress the image into a low-dimensional discrete latent space. After that, the AR models learn from low-dimensional discrete latent space directly, which greatly reduces the time complexity and improves the performance. Liu \textit{et al.} \cite{liu2021conditional} and Iashin and Rahtu \cite{iashin2021taming} also apply the similar idea to generate sound, and achieve good generation performance. \\
% However, they still have weaknesses of unidirectional bias and accumulated prediction errors due to the limitation of AR models.\\
%which first use VQ-VAE \cite{van2017neural} to compress the mel-spectrogram into compact discrete latent space and then AR models are used to model these compressed features
\noindent \textbf{Diffusion Probabilistic Models}
Diffusion generative models were first proposed in \cite{sohl2015deep} and achieved strong results on image generation \cite{dhariwal2021diffusion,gu2021vector,nichol2021glide,esser2021imagebart} and speech synthesis \cite{kong2020diffwave,jeong2021diff,popov2021grad,lee2021priorgrad}.
% Roughly speaking, the diffusion model samples the data distribution by reversing a forward diffusion process that gradually corrupts the input via a fixed Markov chain. The forward process yields a sequence of increasingly noisy latent variables of the same dimensionality as the input, producing pure noise after a fixed number of timesteps. Starting from this noise result, the reverse process gradually denoises the latent variables towards the desired data distribution by learning the conditional transit distribution. We can divide diffusion models into two different kinds according to the latent spaces is continuous or discrete. For continuous diffusion models, we usually add Gaussian noise into the original data on the forward process, and the input becomes noise after several steps. The reverse process aims to gradually recover the original data. 
% Most previous works only considered continuous diffusion models on the raw image pixels or raw speech signals. 
Diffusion models with discrete state spaces were first introduced by Sohl-Dickstein \textit{et al.} \cite{sohl2015deep}, who considered a diffusion process over binary random variables. Hoogeboom \textit{et al.} \cite{hoogeboom2021argmax} extended the model to categorical random variables with transition matrices characterized by uniform transition probabilities. Jacob \textit{et al.} propose {\color{black}{Discrete Denoising Diffusion Probabilistic Models}} (D3PMs) \cite{austin2021structured}, which further improve and extend discrete diffusion models by using a more structured categorical corruption process to corrupt the forward process. D3PMs \cite{austin2021structured} and VQ-Diffusion \cite{gu2021vector} have applied discrete diffusion models to image generation.

\section{Proposed text-to-sound framework} \label{AR sound generation model}
The overall architecture of the proposed text-to-sound framework is demonstrated in Figure \ref{fig:1} (a), which consists of four parts including a text encoder, a VQ-VAE, a token-decoder, and a vocoder. The detailed design of each part will be introduced in this section.
% Figure \ref{fig:1} (a) shows the overall framework of the autoregressive baseline. In this section, we will introduce how to build an autoregressive text-to-sound generation model, which includes four main parts: text encoder, VQ-VAE, AR decoder, and vocoder.
\subsection{Text encoder} \label{text encoder}
The first step in the text-to-sound generation task is designing a good text encoder to extract the sound event information from the context while other information should be excluded. In this study, we employed the pretrained BERT \cite{devlin2018bert} and the text encoder of a pretrained Contrastive Language-Image Pre-Training (CLIP) model \cite{radford2021learning} to extract the text features (a vector to represent the contextual information). Our experiments indicated that using CLIP model can bring better generation performance. {\color{black} One possible explanation is that CLIP is trained by contrastive learning between the representations of images and text, the use of multi-modality information may make the text representations computed from CLIP contain more semantics related to sound events \textit{e.g.}, "dog barks and birds sing".}  Note that the text encoder is fixed in our training process.
% We conjecture that the CLIP model is more suitable for text-to-sound generation tasks for the reason that CLIP is trained by contrastive learning between the representations of images and text, which may make the text representations include more the concept of semantic, \textit{e.g.}, dog barks and birds sing.
% Specifically, we first use BPE-encoding \cite{sennrich2015neural} to transfer the text into a fixed number of token sequences (we set the sequence length as 77) and then extract the text feature representations using the CLIP model. We fix the text encoder in our training process. \\
% the learned relationship between the images and the text is also useful for the text and sound finding that the latter brings better performance. We conjecture that the CLIP model is more suitable for text-to-sound generation tasks for the reason that CLIP is trained by contrastive learning between the representations of images and text, which makes the text representations have more the concept of semantic, \textit{e.g.} A picture of the dog, a caption about the dog, and a dog bark sound all point out the same concept.
\subsection{Learning Discrete Latent Space of Mel-spectrograms Via VQ-VAE} \label{sec:vqvae}
In this part, we introduce the vector quantized variational autoencoder (VQ-VAE) \cite{van2017neural} to simplify the process of decoder generates the mel-spectrograms. 

Most of the text-to-speech (TTS) methods \cite{tan2021survey,kong2020diffwave,jeong2021diff} directly learn the mapping from text to wave samples or raw spectrogram pixels for the reason that the synthesized speech's content {\color{black}{relies on}} the words of text. Unlike TTS, there is no direct correspondence between text and sound in the sound generation task. To this end, the text-to-sound task needs to extract the event information from the text input and then generate the corresponding events. Considering a sound may consist of multiple events and each event has its own unique characteristics. We propose to use the VQ-VAE to learn a codebook to encode the characteristic of events, and then generate the mel-spectrogram based on the codebook {\color{black}{following Liu \textit{et al.} \cite{liu2021conditional} and Iashin and Rahtu \cite{iashin2021taming}}}. 
% \sout{Liu \textit{et al.} \cite{liu2021conditional} and Iashin and Rahtu \cite{iashin2021taming} also employ VQ-VAE to generate sound.} 

As Figure \ref{fig:gan} shows, a mel-spectrogram can be approximated by a sequence of mel-spectrogram tokens. Thus, the mel-spectrogram generation problem transfers to predicting a sequence of tokens. In the following, we will introduce the details of VQ-VAE.
% Transformer architectures have shown great promise in
% image generation \cite{chen2020generative,esser2021taming} and sound generation \cite{liu2021conditional,iashin2021taming} due to their outstanding expressivity. 
% In this work, we follow the SOTA backbone of the video-to-sound generation task, leveraging the transformer to learn the mapping from text to sound. However, if we directly operate on wave samples or raw spectrogram pixels, it leads to high computation costs due to the quadratic nature of the dot-product attention in the transformer. Inspired by the success of sound generation \cite{iashin2021taming,liu2021conditional}, we employ the solution from the computer vision field \cite{razavi2019generating,van2017neural}, which proposes to represent an image by discrete image tokens with reduced sequence length. We use vector quantized variational autoencoder (VQ-VAE) \cite{van2017neural} model to compress the spectrograms into a small size of discrete spectrogram tokens.

% $n_z$ is the dimension of the codebook entries and 
VQ-VAE is trained to approximate an input using a compressed intermediate representation, retrieved from a discrete codebook. VQ-VAE consists of a spectrogram encoder $E_{vq}$, a spectrogram decoder $D_s$ and a codebook $\boldsymbol{Z}=\{ \boldsymbol{z}_k\}^K_{k=1} \in \mathbb{R}^{K \times n_z}$ containing a finite number of embedding vectors, where $K$ is the size of the codebook and $n_z$ is the dimension of codes. Given a mel-spectrogram $\boldsymbol{s} \in \mathbb{R}^{F \times L}$, the input $\boldsymbol{s}$ is firstly encoded into a small-scale representation (encoder $E_{vq}$ consists of multiple convolution and pooling layers) $\boldsymbol{\hat{z}}=E_{vq}(\boldsymbol{s}) \in \mathbb{R}^{F^{\prime} \times L^{\prime} \times n_z}$, where $F^{\prime}$, $L^{\prime}$ and $n_z$ represent the {\color{black}down-sampled frequency, the time dimension and the feature dimension.} 
% \sout{Then we use a spatial-wise quantizer $Q(.)$ which maps each spatial feature $\boldsymbol{\hat{z}}_{ij}$ into its closest codebook entry $\boldsymbol{z}_k$ to obtain a spatial collection of spectrogram tokens $\boldsymbol{z}_q$}
{\color{black}{Then we use a vector quantizer $Q(\cdot)$ which maps each time-frequency vector $\boldsymbol{\hat{z}}_{ij}\in\mathbb{R}^{n_z}$ into its closest codebook entry $\boldsymbol{z}_k$ to obtain a discrete spectrogram token sequence $\boldsymbol{x}\in\mathbb{Z}^{F'\times L'}$} as follows:}
\begin{equation}\label{codebook}
    \boldsymbol{x} = Q(\boldsymbol{\hat{z}}) := \big{(}\mathop{\arg\min}\limits_{\boldsymbol{z}_k \in \boldsymbol{Z}}||\boldsymbol{\hat{z}}_{ij}-\boldsymbol{ z}_k||_2^2 \ \mbox{for all} \ (i,j) \ \mbox{in} \ (F^{\prime}, L^{\prime})\big{)}
\end{equation}
%\in \mathbb{R}^{F^{\prime} \times T^{\prime} \times n_z}
The mel-spectrogram can be {\color{black}approximately} reconstructed via the codebook $\boldsymbol{Z}$ and the spectrogram decoder i.e., $\hat{\boldsymbol{s}}=D_s(Q(\boldsymbol{\hat{z}}))$. Note that the spectrogram tokens are quantized latent variables in the sense that they take discrete values. The encoder $E_{vq}$, the spectrogram decoder $D_s$, and the codebook $\boldsymbol{Z}$ can be trained {\color{black} in an end-to-end manner} via the following loss function:
% \begin{equation}\label{vqvae loss}
%     \mathcal{L}_{\mathit{VQVAE}} = ||\boldsymbol{s}-\boldsymbol{\hat{s}}||_1 +||sg[E_{vq}(\boldsymbol{s})]-\boldsymbol{z}_q||_2^2 + ||sg[\boldsymbol{z}_q]-E_{vq}(\boldsymbol{s})||_2^2
% \end{equation}
\begin{align}\label{vqvae loss}
    \mathcal{L}_{\mathit{VQVAE}} =& ||\boldsymbol{s}-\boldsymbol{\hat{s}}||_1 +||\mathcal{SG}[E_{vq}(\boldsymbol{s})]-\boldsymbol{x}||_2^2 \\ +& ||\mathcal{SG}[\boldsymbol{x}]-E_{vq}(\boldsymbol{s})||_2^2 \noindent
\end{align}
where $\mathcal{SG}[\cdot]$ is the stop-gradient operation that acts as an identity during the forward pass but has zero gradients at the backward pass. To preserve the reconstruction quality when upsampled from a smaller-scale representation, we follow the setting of VQGAN \cite{esser2021taming}, which adds a patch-based adversarial loss \cite{isola2017image} to the final training loss
\begin{equation}\label{SpecVQVAE loss}
    \mathcal{L}_{\mathit{f}} =  \mathcal{L}_{\mathit{VQVAE}} + \lambda_d(\log(D(\boldsymbol{s})) + \log(1-D(\boldsymbol{\hat{s}})))
\end{equation}
% Suppose the size of the VQ-VAE codebook is $K$, for any spectrogram token $x_i \in \{1,2, ..., K\}$.
where $D$ is a discriminator (it consists of several convolution layers), which aims to distinguish the mel-spectrogram is original or reconstructed. $\lambda_d$ is a hyper-parameter to control the weight of adversarial loss.
\subsection{Token-decoder} \label{sec:autoregressive model}
% \sout{The token-decoder in our framework is proposed to transfer the text features to the quantized features obtained from VQ-VAE (mel-spectrogram tokens). An autoregressive token-decoder is first investigated in this paper.}
The token-decoder in our framework is used to transfer the text features into the discrete mel-spectrogram token sequence. An autoregressive token-decoder is first investigated in this paper.

% \sout{Specifically, given the text-audio pairs, we  use the text encoder to extract text features $\boldsymbol{y}$ from the text description as the input of the token decoder.}
{\color{black}Specifically, given the text-audio pairs, the inputs of the token encoder are extracted from the text description with the text encoder.}
% \sout{After that, we obtain the discrete mel-spectrogram tokens $\boldsymbol{x} \in \mathbb{Z}^N$ from the mel-specrogram of the audio with a pre-trained VQ-VAE, where $N=F^{\prime} \times L^{\prime}$ represents the sequence length of tokens.We scan down-then-across the 2D matrix ($F^{\prime} \times L^{\prime}$) to a sequence.}
{\color{black} Following \cite{iashin2021taming}, the discrete mel-spectrogram tokens $\boldsymbol{x}\in\mathbb{Z}^{F'\times L'}$ obtained from a pretrained VQ-VAE are first reshaped to a token sequence $\hat{\boldsymbol{x}}\in\mathbb{Z}^{1 \times F{\prime}L{\prime}}$ and then used as the training target.}
By using the AR token-decoder, the decoding process can be viewed as an autoregressive next-token prediction: Given tokens $\hat{x}_{<i}$, the decoder learns to predict the distribution of possible next tokens, \textit{i.e.}, $p(\hat{x}_i|\hat{\boldsymbol{x}}_{<i},\boldsymbol{y})$ to compute the likelihood of the full representation as $p(\hat{\boldsymbol{x}}|\boldsymbol{y})=\prod_{i}p(\hat{x}_i|\hat{\boldsymbol{x}}_{<i},\boldsymbol{y})$. 
The decoder is trained with a cross-entropy (CE) loss, comparing  {\color{black}the probabilities of the predicted} mel-spectrogram tokens to those obtained from the ground truth. 
Due to the wrongly predicted results of previous steps influencing the current step, ``teacher-forcing'' strategy \cite{esser2021imagebart} is used to guarantee the stability of training. 
{\color{black}{The ``teacher-forcing'' strategy uses ground truth tokens as previous step prediction results.}} 
% \sout{Lastly, we use the decoder of VQ-VAE ($G$) to transform the predicted mel-spectrogram tokens into the mel-spectrogram.} 
{\color{black}After the token decoding process, the mel-spectrogram can be computed by the pretrained VQ-VAE spectrogram decoder.}
{\color{black}In the inference stage, we can set $L^{\prime}$ to determine the duration of the generated sound.} 
% \sout{In practice, $N=F^{\prime} \times L^{\prime}$ and $F^{\prime}$ is fixed, thus we can set different $L^{\prime}$ to determine the duration of the generated sound.}

% \sout{In this study, we speculate the AR decoder suffers from the unnatural bias and accumulation of errors problems based on AR model always predicts tokens in order.}
{\color{black}In the AR token-decoder, the adoption of the ``teacher-forcing`` training strategy can cause a mismatch between model training and inference. During training, we use the previous ground truth tokens to predict the current token, while during inference, we use the predicted tokens. The accuracy of the predicted tokens can be affected by the ``accumulated errors'' in previous predicted tokens in inference \cite{schmidt2019generalization}. Such mismatches can cause the model's performance to be sub-optimal.}
{\color{black} In addition, the prediction of the current token only depends on the previous tokens in the AR decoder, which ignores the future context information. Such ``unidirectional bias'' can also lead to suboptimal model performance.}
To this end, a non-autoregressive token-decoder based on a discrete diffusion model is proposed. (Details will be given in Section \ref{sec:diff-decoder}.)

\subsection{Vocoder}
The vocoder aims at transforming the generated mel-spectrogram into waveform $\hat{w}$. This type of vocoder is a hot research topic. Griffin-Lim \cite{griffin1984signal}, WaveNet \cite{oord2016wavenet}, MelGAN \cite{kumar2019melgan}, and HiFi-GAN \cite{kong2020hifi} are very popular vocoders for speech synthesis task. The Griffin-Lim method is a classic signal processing method that is very fast and easy to implement. However, Griffin-Lim produces low-fidelity results when operating on mel-spectrograms \cite{iashin2021taming}. WaveNet provides high-quality results but remains relatively slow in generation time. In this study, considering its generation efficiency and quality, we employ MelGAN which is a non-autoregressive approach to reconstructing the waveform. MelGAN has been widely used in speech synthesis fields. However, many pre-trained MelGAN models are trained on speech or music data, so they are not suitable for environmental sound generation. We train a MelGAN on a large-scale audio event dataset (AudioSet) \cite{gemmeke2017audio}, which contains 527 unique sound events.
\section{Diffusion-based Decoder} \label{sec:diff-decoder}
In this section, we introduce our proposed non-autoregressive token-decoder based on discrete diffusion model, named Diffsound. As discussed in Section \ref{sec:autoregressive model}, Diffsound model is proposed to address the unnatural bias and accumulation of errors issues in AR decoders. In the following, we first introduce the diffusion models. Then we discuss the details of the training and inference of the Diffsound model. Lastly, we discuss how to use the pre-training strategy to further improve the performance of the Diffsound model.
% In previous section, we introduce how to use autoregressive manner to generate sound. However, autoregressive model requires a forward pass of the network to predict each token, which consumes an inordinate amount of time when we want to generate long-duration sound. Furthermore, we use the ``teacher-forcing'' strategy in the training stage, which will lead to error accumulation due to the mistakes in the earlier sampling, because we cannot get the ground truth in inference stage. Thus we propose use discrete diffusion model to generate desired tokens simultaneously.
\subsection{Diffusion Probabilistic Models} \label{sec:dpm}
Diffusion probabilistic models (diffusion models for short) \cite{sohl2015deep} have proved to be a powerful generation model in the image and speech fields \cite{dhariwal2021diffusion,gu2021vector}. In this section, we briefly introduce some basic principles of the diffusion models.
\subsubsection{Vanilla Diffusion Model}
A diffusion model consists of two processes: the \textit{forward} process with steps $t \in \{0, 1, 2, ..., T \}$ and the \textit{reverse} process $t \in \{T, T-1, ..., 1, 0 \}$. The forward process corrupts the original data $\boldsymbol{x}_0$ into a noisy latent variable $\boldsymbol{x}_T$ which belongs to a stationary distribution (\textit{e.g.}, Gaussian distribution), and the reverse process learns to recover the original data $\boldsymbol{x}_0$ from $\boldsymbol{x}_T$.\\
\textbf{Forward process}
Given the audio data $\boldsymbol{x}_0$, the forward process aims to corrupt the data $\boldsymbol{x}_0 \sim q(\boldsymbol{x}_0)$ into a sequence of increasingly noisy latent variables $\boldsymbol{x}_{1:T}=\boldsymbol{x}_1, \boldsymbol{x}_2, ..., \boldsymbol{x}_T$. Each of noisy latent variables $\boldsymbol{x}_t$ has the {\color{black}same dimensionality as} $\boldsymbol{x}_0$. The forward process from data $\boldsymbol{x}_0$ to the
variable $\boldsymbol{x}_T$ can be formulated as a fixed Markov chain
\begin{equation}\label{forward process}
    q(\boldsymbol{x}_{1:T}|\boldsymbol{x}_0)=\prod_{t=1}^{T}q(\boldsymbol{x}_t|\boldsymbol{x}_{t-1})
\end{equation}
Following \cite{sohl2015deep}, Gaussian noise is selected in each step, so that the conditional probability distribution can be  $q(\boldsymbol{x}_t|\boldsymbol{x}_{t-1})=\mathcal{N}(\boldsymbol{x}_t;\sqrt{1-\beta_{t}}\boldsymbol{x}_{t-1}, \beta_t \boldsymbol{I})$, where $\beta_{t}$ is a small positive constant. {\color{black}{According to the pre-defined schedule $\beta_{1}, \beta_{2}, ..., \beta_{T}$ (detials are given in Section \ref{experimental setup}), the forward process gradually converts original $\boldsymbol{x}_0$ to a latent variable with an isotropic Gaussian distribution of $p(\boldsymbol{x}_T) =\mathcal{N}(\boldsymbol{0}, \boldsymbol{I})$ when $T$ is enough large \cite{sohl2015deep}.}} 
Based on the properties of Markov chain \cite{ho2020denoising}, the probability distribution $q(\boldsymbol{x}_t|\boldsymbol{x}_0)$ can be written as
\begin{equation}\label{forward process}
    q(\boldsymbol{x}_{t}|\boldsymbol{x}_0)= \mathcal{N}(\boldsymbol{x}_t;\sqrt{\overline{\alpha}_t}\boldsymbol{x}_{0},(1-\overline{\alpha}_t) \boldsymbol{I})
\end{equation}
where $\alpha_t=1-\beta_t$ and $\overline{\alpha}_t= \prod_{s=1}^{t} \alpha_s$. \\
\textbf{Reverse process}
The reverse process converts the latent variable $\boldsymbol{x}_T \sim \mathcal{N}(\boldsymbol{0},\boldsymbol{I}) $ into $\boldsymbol{x}_0$, whose joint probability follows:
\begin{equation}\label{forward process}
   p_{\theta}(\boldsymbol{x}_{0:T})=p(\boldsymbol{x}_T) \prod_{t=1}^{T} p_{\theta}(\boldsymbol{x}_{t-1}|\boldsymbol{x}_t)
\end{equation}
where $p_{\theta}(\cdot)$ is the distribution of the reverse process with learnable parameters $\theta$. The posterior $q(\boldsymbol{x}_{t-1}|\boldsymbol{x}_t,\boldsymbol{x}_0)$ can be derived according to Bayes formula as follows:
{\color{black}
\begin{equation}\label{bayespost}
\begin{aligned}
    q(\boldsymbol{x}_{t-1}|\boldsymbol{x}_{t},\boldsymbol{x}_{0})= \frac{q(\boldsymbol{x}_t|\boldsymbol{x}_{t-1},\boldsymbol{x}_0)q(\boldsymbol{x}_{t-1}|\boldsymbol{x}_0)}{q(\boldsymbol{x}_t|\boldsymbol{x}_0)} 
\end{aligned}
\end{equation}
}
In order to optimize the generative model $p_{\theta}(\boldsymbol{x}_0)$ to fit the data distribution $q(\boldsymbol{x}_0)$, one typically optimizes a variational upper bound on the negative log-likelihood {\color{black}{\cite{ho2020denoising}}}:
\begin{equation}\label{vqvae loss}
\begin{aligned}
    \mathcal{L}_{\mathit{vb}} =  \mathbb{E}_{q(\boldsymbol{x}_0)} \Big{[}   D_{KL} [ q(\boldsymbol{x}_T|\boldsymbol{x}_0)||p(\boldsymbol{x}_T) ]   + \\  
    \sum_{t=1}^{T}   \mathbb{E}_{q(\boldsymbol{x}_t|\boldsymbol{x}_0)} \big{[}D_{KL}[q(\boldsymbol{x}_{t-1}|\boldsymbol{x}_t,\boldsymbol{x}_0)||p_{\theta}(\boldsymbol{x}_{t-1}|\boldsymbol{x}_t)] \big{]} \Big{]}. 
\end{aligned}
\end{equation}
\subsubsection{Discrete Diffusion model} \label{bkg:ddm}
One limitation of vanilla diffusion model is that, for original data $\boldsymbol{x}_0$ in discrete space, \textit{e.g.}, for any element $x_0^r$ in $\boldsymbol{x}_0$, $x_0^r \in \{1, 2, ..., P\}$, we cannot  corrupt the $\boldsymbol{x}_0$ by adding Gaussian noise in the forward process since the range of $x_0^r$ belongs to $P$ different discrete values. To solve this issue, discrete diffusion model \cite{sohl2015deep,austin2021structured} is proposed.
In discrete diffusion model, a transition probability matrix is defined to indicate how $\boldsymbol{x}_0$ transits to $\boldsymbol{x}_t$ for each step of forward process. Assuming that $\boldsymbol{x}_0 \in \mathbb{Z}^N$ and $x_{0}^{k} \in \{1, 2, ..., P\}$. It is worth noting that $\boldsymbol{x}_0$ denotes a vector with $N$ scalar elements. {\color{black}To better illustrate the transition probability matrix, we will use $x_0$ to denote any one scalar element in $\boldsymbol{x}_0$ in the following.} The matrices $[\boldsymbol{Q}_t]_{ij}=q(x_t=i|x_{t-1}=j) \in \mathbb{R}^{P \times P}$ defines the probabilities that $x_{t-1}$ transits to $x_t$. Then the forward process for the whole token sequence can be written as:
\begin{equation}\label{discrete forward process}
     q(x_t|x_{t-1})=\boldsymbol{c}^\top(x_t) \boldsymbol{Q}_t \boldsymbol{c}(x_{t-1})
\end{equation}
where $\boldsymbol{c}(\dot)$ denotes a function that can transfer a scalar element into a one-hot column vector. 
The categorical distribution over $x_t$ is given by the vector $\boldsymbol{Q}_t \boldsymbol{c}(x_{t-1})$. 
{\color{black} Due to the property of Markov chain}, one can marginalize out the intermediate steps and derive the probability of $x_t$ at arbitrary timestep directly from $x_0$ as follows:
\begin{equation}\label{q(x_t|x_0)}
    q(x_t|x_{0})=\boldsymbol{c}^\top(x_t) \overline{\boldsymbol{Q}}_t \boldsymbol{c}(x_{0}), \mathit{with} \ \overline{\boldsymbol{Q}}_t= \boldsymbol{Q}_t \dots \boldsymbol{Q}_1
\end{equation}
Therefore, $q(x_{t-1}|x_{t},x_{0})$ can be computed based on Eq.\ref{q(x_t|x_0)}:
\begin{equation}\label{post}
\begin{aligned}
    q(x_{t-1}|x_{t},x_{0})= \frac{q(x_t|x_{t-1},x_0)q(x_{t-1}|x_0)}{q(x_t|x_0)} \\
    = \frac{\big{(}\boldsymbol{c}^\top(x_t) \boldsymbol{Q}_t \boldsymbol{c}(x_{t-1}) \big{)} \big{(}\boldsymbol{c}^\top(x_{t-1}) \overline{\boldsymbol{Q}}_{t-1} \boldsymbol{c}(x_0) \big{)}}{\boldsymbol{c}^\top(x_t) \overline{\boldsymbol{Q}}_t \boldsymbol{c}(x_0)}
\end{aligned}
\end{equation}
\subsection{Non-autoregressive Mel-spectrograms Generation Via Diffsound}
% Given the text-audio pairs, we can obtain the discrete spectrogram tokens $\boldsymbol{x} \in  \mathbb{Z}^N $ with a pre-trained VQ-VAE. We suppose the size of the VQ-VAE codebook is $K$, so for any spectrogram token $x_i \in \{1,2, ..., K\}$.
% Similar to the AR decoder, we use the diffusion-based decoder to generate the spectrogram tokens and then use the decoder of VQ-VAE to transform the tokens into mel-spectrogram.
In contrast to the AR token-decoder, which predicts the mel-spectrogram tokens one by one, we expect the Diffsound model to predict all of the tokens in a non-AR manner. {\color{black}{Specifically, the Diffsound model can predict all of the tokens simultaneously, then refine the predicted results in the following steps so that the best results can be obtained through iterations.}} In other words, we expect the predicted results can be improved through $T$-step iterations. {\color{black}{In contrast, AR decoder needs $N$ steps to get results, where $N$ denotes the number of tokens (in practice, $N \gg T$).}} {\color{black}{The Diffsound model can make use of the contextual information of all tokens and revise any token in each step. We speculate that it effectively diminishes the unnatural bias and the accumulated prediction error problems.}} To realize the process, we adapt the idea from discrete diffusion model \cite{sohl2015deep,austin2021structured,gu2021vector}, which designs a strategy to corrupt the original mel-spectrogram token sequence $\boldsymbol{x}_0 \sim q(\boldsymbol{x}_0) $ into a totally meaningless sequence $\boldsymbol{x}_T \sim p(\boldsymbol{x}_T)$ in $T$ steps, and then let network learn to recover the original sequence $\boldsymbol{x}_0$ based on the text information in $T$ steps. Figure \ref{fig:1} (c) shows an example of the forward and reverse processes. In the inference stage, the reverse process is used to help generate mel-spectrogram tokens. We randomly sample a token sequence $\boldsymbol{x}_T$ from $p(\boldsymbol{x}_T)$, and then let the network predict a new mel-spectrogram token sequence based on $\boldsymbol{x}_T$ and the text features. According to the previous description in Section \ref{sec:dpm}, the key point of training a discrete diffusion model is to design a suitable strategy to pre-define Markov transition matrices $\boldsymbol{Q}_t$.
% and corresponding stationary distributions $p(\boldsymbol{x}_T)$. 

As discussed in Section \ref{sec:vqvae}, the codebook of VQ-VAE encodes the time-frequency features of sound events. According to this property, we propose three strategies to corrupt the mel-spectrogram tokens. Firstly, changing the context by randomly replacing the original token. Secondly, masking the context by introducing an extra mask token. {\color{black}{However, we find that changing the context by randomly replacing tokens results in the reverse process being hard to learn. We speculate that one of the reasons is that a mel-spectrogram token may be replaced with a completely unrelated category token, \textit{e.g.}, a token representing the dog barking may be replaced by a token representing the man speaking. Furthermore, we conjecture that there is a context relationship between adjacent tokens; if we only use the mask token, the model may tend to focus on the mask token and ignore the context relationship.}} Thus, we propose to combine the changing and masking context strategies. We define three transition matrices according to the three strategies, respectively: Uniform transition matrix, mask transition matrix, and mask and uniform transition matrix.

In the following, we first introduce the three transition matrices. Then we discuss the noise schedule strategy and loss function. Lastly, we introduce the reparameterization trick and fast inference strategy.  \\
% \subsubsection{Choice of Markov transition matrices for the forward process}
% In this part, we discuss how to use discrete diffusion model to generate spectrograms. \\
 \noindent \textbf{Uniform transition matrix} Uniform transition matrix was first proposed by Sohl-Dickstein \textit{et al.} \cite{sohl2015deep} for binary random variables, \textit{e.g.}, variable zero can transfer to one or zero with a uniform distribution. Hoogeboom \textit{et al.} \cite{hoogeboom2021argmax} later extended this to categorical variables. The core idea is that each variable can transfer to all the pre-defined categories with a uniform distribution. In practice, the transition matrix $\boldsymbol{Q}_t \in \mathbb{R}^{K \times K}$ can be defined as
 \begin{equation} \label{uniform transition mat}
 \boldsymbol{Q}_t =
 \begin{bmatrix}
    \alpha_t + \beta_t & \beta_t & \cdots & \beta_t \\
    \beta_t &  \alpha_t + \beta_t & \cdots & \beta_t \\
    \vdots  &   \vdots            &  \ddots  &    \vdots \\
    \beta_t & \beta_t  &  \cdots & \alpha_t + \beta_t
\end{bmatrix}
\end{equation}
% Since this transition matrix is doubly stochastic with strictly positive entries
where $\beta_t \in [0,1]$ and $\alpha_t=1-K\beta_t$. This transition matrix denotes that each token has a probability of $K\beta_t$ to be resampled uniformly over all the K categories, while with a probability of $\alpha_t$ to remain the previous value at the present step. 
As Section \ref{bkg:ddm} described, we could calculate $q(x_t|x_0)$ according to formula (\ref{q(x_t|x_0)}), $q(x_t|x_0)=\boldsymbol{c}^\top(x_t) \overline{\boldsymbol{Q}}_t \boldsymbol{c}(x_{0})$. However, when the number of categories $K$ and time step $T$ is too large, it can quickly become impractical to store all of the transition matrices $\boldsymbol{Q}_t$ in memory, as the memory usage grows like $O(K^2T)$. Inspired by \cite{gu2021vector}, we find that it is unnecessary to store all of the transition matrices. Instead we only store all of $\overline{\alpha}_t$ and $\overline{\beta}_t$ in advance, because we can calculate $q(x_t|x_0)$ according to following formula:
\begin{equation}\label{formula:quick cal uniform matrix}
   \overline{\boldsymbol{Q}}_t \boldsymbol{c}(x_0) = \overline{\alpha}_t \boldsymbol{c}(x_0) + \overline{\beta}_t.
\end{equation}
where $\overline{\alpha}_t=\prod_{t=1}^{t}\alpha_t$, $\overline{\beta}_t=(1- \overline{\alpha}_t)/K$.
When $T$ is enough large, $\overline{\alpha}_t$ is close to 0. Thus, we can derive the stationary distribution $p(\boldsymbol{x}_T)$ as:
\begin{equation}\label{formula:uniform transition matrices}
   p(\boldsymbol{x}_T)=[\overline{\beta}_T, \overline{\beta}_T, \cdots, \overline{\beta}_T]
\end{equation}
where $\overline{\beta}_T=1/K$.
The Proof can be found in Appendix \ref{appendix1}.
% Austin \textit{at.al} \cite{austin2021structured} propose two approaches (low-rank corruption and matrix exponentials ) to scale $\overline{\boldsymbol{Q}}_t$, which  
\begin{algorithm}[t]
\caption{Training of the Diffsound model.}
\label{alg:PA1}
\begin{algorithmic}[1]
\REQUIRE ~~\\
    A transition matrix $\boldsymbol{Q}_t$, the number of timesteps $T$, network parameters $\theta$, training epoch $N$, text-audio dataset $\boldsymbol{D}$, the encoder of VQ-VAE $E_{vq}$.
    \FOR{$i=1$ to $N$}
    \FOR{$(\mbox{text},\mbox{audio})$ in $
    \boldsymbol{D}$}
    \STATE $\boldsymbol{s} = \mbox{get\_mel\_spectrogram(audio)}$;
    \STATE $\boldsymbol{x}_0 = E_{vq}(\boldsymbol{s})$, $\boldsymbol{y}=$TextEncoder($\mbox{text}$);
    \STATE sample $t$ from Uniform($1, 2, 3, ..., T$);
    \STATE sample $\boldsymbol{x}_t$ from $q(\boldsymbol{x}_t|\boldsymbol{x}_0)$ based on formula (\ref{formula:quick cal uniform and mask matrix});
    \STATE estimate $p_{\theta}(\boldsymbol{x}_{t-1}|\boldsymbol{x}_t,\boldsymbol{y})$;
    \STATE calculate loss according to formula (\ref{vlb loss}-\ref{diffu final loss});
    \STATE update network $\theta$;
    \ENDFOR
    \ENDFOR
\RETURN network $\theta$.
\end{algorithmic}
\end{algorithm}

\noindent\textbf{Mask transition matrix} Previous work \cite{hoogeboom2021argmax,gu2021vector} proposed introducing an additional absorbing state, such that each token either stays the same or transitions to the absorbing state. In this study, we add an additional token [MASK] into the codebook (the index is $K+1$) to represent the absorbing state, thus the model is asked to recover the original tokens from the mask tokens. The mask transition matrix $\boldsymbol{Q}_t \in \mathbb{R}^{(K+1) \times (K+1)}$ is
\begin{equation} \label{masked transition mat}
 \boldsymbol{Q}_t =
 \begin{bmatrix}
    \beta_t  & 0 & 0 & \cdots & 0 \\
    0 &  \beta_t & 0  & \cdots & 0  \\
    0 & 0 & \beta_t  & \cdots  & 0  \\
    \vdots  &   \vdots     &  \vdots      &  \ddots  &    \vdots \\
    \gamma_t & \gamma_t & \gamma_t   &  \cdots & 1
\end{bmatrix}
\end{equation}
% not uniform but to the point-mass distribution on [MASK] token
% This matrix does not impose particular relationships between categories but allows corrupted tokens to be distinguished from original ones.
\noindent where $\gamma_t \in [0,1]$. The mask transition matrix denotes that each token either stays the same with probability $\beta_t =(1-\gamma_t)$ or transitions to an additional token [MASK] with probability $\gamma_t$. Similarly, we only store all of $\overline{\gamma}_t$ and $\overline{\beta}_t$ in advance, and we calculate $q(x_t|x_0)$ according to following formula:
\begin{equation}\label{formula:quick cal mask matrix}
   \overline{\boldsymbol{Q}}_t \boldsymbol{c}(x_0) = \overline{\beta}_t \boldsymbol{c}(x_0) + \overline{\gamma}_t \boldsymbol{c}(K+1) .
\end{equation}
where $\overline{\gamma}_t= 1- \overline{\beta}_t$, and $\overline{\beta}_t=\prod_{t=1}^{t} \beta_t$.
According to Markov transition formula, when $T$ is enough large, $\overline{\beta}_T$ is close to 0. Thus the stationary distribution $p(\boldsymbol{x}_T)$ can be derived as:
\begin{equation}\label{formula:masked transition matrices}
  p(\boldsymbol{x}_T)=[0, 0 , \cdots, 1].
\end{equation}
The Proof can be found in Appendix \ref{appendix2}. \\
\textbf{Mask and uniform transition matrix} As in the previous discussion, {\color{black}{we speculate that using the uniform transition matrix brings the reverse process is hard to learn. Using the mask transition matrix may make the model tend to focus on the mask token and ignore the context information.}} To combat these problems, a simple idea is to combine mask and uniform transition matrices. Specifically, each token has a probability of $\gamma_t$ to transition to [MASK] token, a probability of $K\beta_t$ be resampled uniformly over all the K categories and a probability of $\alpha_t =  1-K\beta_t-\gamma_t$ to stay the same token. The transition matrices $\boldsymbol{Q}_t \in \mathbb{R}^{(K+1) \times (K+1)}$ is defined as
\begin{equation} \label{masked and uniform transition mat}
 \boldsymbol{Q}_t =
 \begin{bmatrix}
    \alpha_t + \beta_t  & \beta_t & \beta_t & \cdots & 0 \\
    \beta_t &  \alpha_t + \beta_t & \beta_t  & \cdots & 0  \\
    % \beta_t & \beta_t & \alpha_t + \beta_t & \cdots  & 0  \\
    \vdots  &   \vdots     &  \vdots      &  \ddots  &    \vdots \\
    \gamma_t & \gamma_t & \gamma_t   &  \cdots & 1
\end{bmatrix}.
\end{equation}
According to previous discussions, we can calculate $q(x_t|x_0)$ according to following formula:
\begin{equation}\label{formula:quick cal uniform and mask matrix}
   \overline{\boldsymbol{Q}}_t \boldsymbol{c}(x_0) = \overline{\alpha}_t \boldsymbol{c}(x_0) + ( \overline{\gamma}_t - \overline{\beta}_t )\boldsymbol{c}(K+1) + \overline{\beta}_t .
\end{equation} 
where $\overline{\alpha}_t= \prod_{t=1}^{t} \alpha_t$, $\overline{\gamma}_t=1-\prod_{t=1}^{t}(1-\gamma_t)$ and $\overline{\beta}_t=(1-\overline{\alpha}_t - \overline{\gamma}_t)/K$.
% The random token replacement forces the network to understand the context rather than only focusing on the [MASK] tokens. 
Similarly, when $T$ is enough large, $\overline{\alpha}_T$ is close to 0. Thus the stationary distribution $p(\boldsymbol{x}_T)$ can be derived as
\begin{equation}\label{formula:masked transition matrices}
  p(\boldsymbol{x}_T)=[\overline{\beta}_T, \overline{\beta}_T , \cdots, \overline{\gamma}_T]
\end{equation}
where $\overline{\beta}_T=(1-\overline{\gamma}_T)/K$. \\
\textbf{Noise schedule} Noise schedule is used to pre-define the value of transition matrices (pre-define $\overline{\alpha}_t$, $\overline{\beta}_t$, and $\overline{\gamma}_t$ in our study). Many noise schedules have been proposed, such as the linear schedule, the consine schedule \cite{nichol2021improved}, and the mutual-information-based noise schedule \cite{austin2021structured}. In this study, we adapted the linear schedule for all of the experiments. \\
% Note that a linear schedule for $\boldsymbol{Q}_t$ leads to a nonlinear amount of cumulative noise in $\overline{\boldsymbol{Q}}_t$. \\
\begin{algorithm}[t]
\caption{Inference of the Diffsound model.}
\label{alg:PA2}
\begin{algorithmic}[1]
\REQUIRE ~~\\
    Time stride $\Delta_t$, the number of timesteps $T$, input text, the decoder of VQ-VAE $G$, network $\theta$, stationary distribution $p(\boldsymbol{x}_T)$;
    \STATE $t=T$, $\boldsymbol{y}=$TextEncoder(text);
    \STATE sample $\boldsymbol{x}_t$ from $p(\boldsymbol{x}_T)$;
    \WHILE{$t \textgreater 0$}
    \STATE $\boldsymbol{x}_t \leftarrow $ sample from $p_{\theta}(\boldsymbol{x}_{t-\Delta_t}|\boldsymbol{x}_t,\boldsymbol{y})$
    \STATE $t \leftarrow (t-\Delta_t)$
    \ENDWHILE
\RETURN  $G(\boldsymbol{x}_t)$.
\end{algorithmic}
\end{algorithm}
\textbf{Loss function}
Similar to the training objective of a continuous diffusion model (Eq.\ref{vqvae loss}), we also train a network $p_{\theta}(\boldsymbol{x}_{t-1}|\boldsymbol{x}_{t},\boldsymbol{y})$ to estimate the posterior transition distribution $q(\boldsymbol{x}_{t-1}|\boldsymbol{x}_t,\boldsymbol{x}_0)$. The network is trained to minimize the variational lower bound (VLB).
\begin{equation}\label{vlb loss}
\begin{aligned}
    \mathcal{L}_{\mathit{vlb}} &= \sum_{t=1}^{T-1}  \big{[}D_{KL}[q(\boldsymbol{x}_{t-1}|\boldsymbol{x}_t,\boldsymbol{x}_0)||p_{\theta}(\boldsymbol{x}_{t-1}|\boldsymbol{x}_t,\boldsymbol{y})] \big{]} \\
    &+ D_{KL}(q(\boldsymbol{x}_T|\boldsymbol{x}_0)||p(\boldsymbol{x}_T))  
\end{aligned}
\end{equation}
% \begin{equation}\label{vlb loss}
% \begin{aligned}
%     & \mathcal{L}_{\mathit{vlb}} = \mathcal{L}_{0} +  \mathcal{L}_{1} + \dots +  \mathcal{L}_{T-1} +  \mathcal{L}_{T} \\  
%   & \mathcal{L}_{0} = -log p_{\theta}(\boldsymbol{x}_0|\boldsymbol{x}_1,\boldsymbol{y}),\\
%   & \mathcal{L}_{t-1} =  D_{KL}[q(\boldsymbol{x}_{t-1}|\boldsymbol{x}_t,\boldsymbol{x}_0)||p_{\theta}(\boldsymbol{x}_{t-1}|\boldsymbol{x}_t,\boldsymbol{y})] \\
%   & \mathcal{L}_{T}=D_{KL}(q(\boldsymbol{x}_T|\boldsymbol{x}_0)||p(\boldsymbol{x}_T)) 
% \end{aligned}
% \end{equation}
where $p(\boldsymbol{x}_T)$ is the stationary distribution, which can be derived in advance. 
Note that we add conditional information $\boldsymbol{y}$ to $p_\theta(\boldsymbol{x}_{t-1} | \boldsymbol{x}_t, \boldsymbol{y})$ in Eq.\ref{vlb loss}. 
The completed training algorithm is summarized in Algorithm \ref{alg:PA1}.\\
\textbf{Reparameterization trick} Recent works \cite{gu2021vector,nichol2021improved} found that approximating some surrogate variables, {\color{black}such as $p_{\theta}(\hat{\boldsymbol{x}}_0|\boldsymbol{x}_t,\boldsymbol{y})$ gives better results comparing with directly predicting the posterior $q(\boldsymbol{x}_{t-1}|\boldsymbol{x}_t,\boldsymbol{x}_0)$.} In this study, we follow the reparameterization trick proposed in \cite{gu2021vector}, which lets the Diffsound model predict the noiseless token distribution $p_{\theta}(\hat{\boldsymbol{x}}_0|\boldsymbol{x}_t,\boldsymbol{y})$ at each reverse step, and then compute $p_{\theta}(\boldsymbol{x}_{t-1}|\boldsymbol{x}_t,\boldsymbol{y})$ according to the following formula:
\begin{equation}\label{Reparameterization}
	p_{\theta}(\boldsymbol{x}_{t-1}|\boldsymbol{x}_t,\boldsymbol{y})=\sum_{\hat{\boldsymbol{x}}_0=1}^{K} q(\boldsymbol{x}_{t-1}|\boldsymbol{x}_t,\hat{\boldsymbol{x}}_0) p_{\theta}(\hat{\boldsymbol{x}}_0|\boldsymbol{x}_t,\boldsymbol{y}).
\end{equation}
Based on the formula (\ref{Reparameterization}), an auxiliary denoising objective loss is introduced, which encourages the network to predict $p_{\theta}(\hat{\boldsymbol{x}}_0|\boldsymbol{x}_t,\boldsymbol{y})$:
\begin{equation}\label{noiseless loss}
  \mathcal{L}_{x_0} = -\log p_{\theta}(\hat{\boldsymbol{x}}_0|\boldsymbol{x}_t,\boldsymbol{y}).
\end{equation}
Experimental results indicate that combining $\mathcal{L}_{x_0}$ and $\mathcal{L}_{vlb}$ could get better performance. Thus our final loss function is defined as:
\begin{equation}\label{diffu final loss}
  \mathcal{L} = \lambda\mathcal{L}_{x_0} + \mathcal{L}_{vlb}
\end{equation}
where $\lambda$ is a hyper-parameter to control the weight of the auxiliary loss $\mathcal{L}_{x_0}$.\\
\textbf{Fast inference strategy} We can see that the inference speed of the Diffsound model is related to the number of timesteps, $T$. By leveraging the reparameterization trick, we can skip some steps in the Diffsound model to achieve a faster inference. Usually, we sample the spectrogram tokens in the chain of $\boldsymbol{x}_T, \boldsymbol{x}_{T-1}, \boldsymbol{x}_{T-2}, ..., \boldsymbol{x}_{0}$. Thus, we can use a larger time stride $\Delta_{t}$ by sampling the spectrogram tokens in the chain of $\boldsymbol{x}_T, \boldsymbol{x}_{T-\Delta_{t}}, \boldsymbol{x}_{T-2\Delta_{t}}, ..., \boldsymbol{x}_{0}$. 
{\color{black}Similar to Eq.\ref{Reparameterization}, with the fast inference strategy,  $p_\theta(\boldsymbol{x}_{t-\Delta_t}|\boldsymbol{x}_t, \boldsymbol{y})$ can be computed as:
\begin{equation}\label{Reparameterization2}
	p_{\theta}(\boldsymbol{x}_{t-\Delta_t}|\boldsymbol{x}_t,\boldsymbol{y})=\sum_{\hat{\boldsymbol{x}}_0=1}^{K} q(\boldsymbol{x}_{t-\Delta_t}|\boldsymbol{x}_t,\hat{\boldsymbol{x}}_0) p_{\theta}(\hat{\boldsymbol{x}}_0|\boldsymbol{x}_t,\boldsymbol{y}).
\end{equation}
{Note that we make sure the last step is $\boldsymbol{x}_{0}$ in our experiments}}. We found this strategy makes the inference stage more efficient, which only causes little decrease to quality. The whole inference algorithm is summarized in Algorithm \ref{alg:PA2}.\\
\begin{algorithm}[t]
\caption{Pre-training the Diffsound model on AudioSet.}
\label{alg:PA3}
\begin{algorithmic}[1]
\REQUIRE ~~\\
    The audio and its corresponding label $\{\boldsymbol{A}, \boldsymbol{L}\}$, the number of training epoch $n$, initial network parameters $\theta$;
    \STATE Count the number of sound events for each audio-label pair;
    \STATE Split $\{\boldsymbol{A}, \boldsymbol{L}\}$ into two subset according the number of events, $\{\boldsymbol{A}_{SES},\boldsymbol{L}_{SES}\}, \{\boldsymbol{A}_{MES},\boldsymbol{L}_{MES}\}$;
    \FOR{$i=1$ to $n$}
    \FOR{$(\mbox{audio},l)$ in $\{\boldsymbol{A}_{SES},\boldsymbol{L}_{SES}\}$}
    \STATE $\mbox{text} =$ MBTG($l$);
    \STATE Train the model $\theta$ according to ($\mbox{text},\mbox{audio}$);
    \ENDFOR
    \ENDFOR
    \FOR{$i=1$ to $2 \cdot n$}
    \FOR{$(\mbox{audio},l)$ in $\{\boldsymbol{A}_{MES},\boldsymbol{L}_{MES}\}$}
    \STATE $\mbox{text} =$ MBTG($l$);
    \STATE Train the model $\theta$ according to ($\mbox{text},\mbox{audio}$);
    \ENDFOR
    \ENDFOR
\RETURN Diffsound $\theta$.
\end{algorithmic}
\end{algorithm}
\subsection{Pre-training Diffsound on AudioSet dataset} 
Recently, image generation has got great success, one of the reasons is that they collect large-scale text-image data, \textit{e.g.}, CogView \cite{ding2021cogview} collects more than 30 million high-quality (Chinese) text-image pairs. However, collecting large-scale text-audio pairs is time-consuming. AudioSet \cite{gemmeke2017audio} is the largest open-source dataset in the audio field, but it only provides the event labels for each audio. To utilize these data, we propose a mask-based text generation method to generate text descriptions according to event labels. \\
% \textbf{Directly using the label information}
\textbf{Mask-based text generation method (MBTG)}
Our method is based on our observation of how humans write text descriptions for audio: Humans first listen to the audio to find out which events happen, and then they add some detailed descriptions to compose a sentence. For example, if the label is ``dog barks, man speaks'', one can generate the text description like that ``a dog is barking when a man is speaking'' or ``a dog barks after a man speaks over''. The first sentence indicates that the events of dog barking and man speaking are simultaneously happening. The latter shows that we first listen to a man speaks, and then a dog barks. Although the keywords are the same, the generated two texts correspond to different audio due to different detailed descriptions. It means that automatically generating text descriptions according to the label information is a tough task. Instead of generating specific text, we propose to use `[MASK]' token to replace the detailed description. We can generate text descriptions like that ``[MASK] [MASK] dog bark [MASK] man speaking [MASK]''. We expect the model to learn the relationship of events rather than directly obtain it from the text description. The generation algorithm is easy to implement. We randomly insert $m \in \{1,2\}$ mask tokens on either side of the label. \\
\noindent\textbf{Curriculum Learning Strategy}
We found that it is easier to generate audio that only includes a single event than to audio that includes multiple events. To help the Diffsound model learn better, we mimic the human learning process {\color{black}{by letting}} the Diffsound model learn from easy samples, and gradually advance to complex samples and knowledge. Thus, a curriculum learning \cite{bengio2009curriculum} strategy is proposed. Specifically, we rank the AudioSet according to the number of events in the audio, and then we split the AudioSet into two subsets: one only includes the audio of a single event (we refer to it as the Single Event Set (SES)), and the other includes the audio of multiple events (we refer to it as the Multiple Event Set (MES)). We first train the Diffsound model on the SES in the first few epochs. After that we train the Diffsound model on the MES. The whole algorithm is summarized in Algorithm \ref{alg:PA3}.  
\section{Dataset and Data Pre-processing}
AudioSet \cite{gemmeke2017audio} and AudioCaps \cite{kim2019audiocaps} dataset are used in our experiments. In the following, we first introduce the AudioSet and AudioCaps datasets, then we discuss the details of data pre-processing.
\subsection{AudioSet} 
An ontology comprising 527 sound classes is used in the large-scale audio dataset known as AudioSet. More than 2 million 10 seconds audio snippets from YouTube videos are included in the AudioSet collection. There are roughly 1.9M audio clips in our training set because some audio clips can no longer be downloaded. Each audio clip may have one or more labels for the presented audio events.
\subsection{AudioCaps} AudioCaps is the largest audio captioning dataset currently available with around 50k audio clips sourced from AudioSet. AudioCaps includes three sets: training, validation, test sets. There are 49256, 494, and 957 audio clips in our training, validation and test sets. Each audio clip in the training set contains one human-annotated caption, while each contains five captions in the validation and test set. We use the AudioCaps training set to train our models. We evaluate our methods on the AudioCaps validation set. 
\subsection{Data pre-processing}
All audio clips in the two datasets are converted to 22.05k Hz and
padded to 10 seconds long. Log mel-spectrograms extracted using a 1024-points Hanning window with 256-points hop size and 80 mel bins
are used as the input features. Finally, we can extract a $860 \times 80$ mel-spectrogram from 10 seconds audio. 
\section{Evaluation Metric} \label{sec:evaluation metrix}
In this study, we investigate Humans Mean Opinion Score (MOS) and objective assessment metrics.
\subsection{MOS} We randomly choose 15 sets of generated sound clips by AR token-decoder and Diffsound model. Each set includes one text description, one real sample, 1-2 generated sounds by AR token-decoder, and 1-2 generated sounds by the Diffsound model. We let 10 people give the grades for these sounds in three aspects: relevance, fidelity, and intelligibility. Note that the test person never knows whether the sound is real or generated. We ask people to give 0-5 grades on the three aspects. Finally, we use the average score of the three aspects as the MOS.
\subsection{Objective Assessment Metrics}
Human evaluation of the performance of the sound generation model is an expensive and tedious procedure. Thus, designing a proper evaluation metric that can measure the gap between generated and real samples is very important for the generation task. In this paper, we employed three objective quality assessment metrics: Fréchet Inception Distance (FID), Kullback-Leibler (KL) divergence, and Audio Caption Loss. We also conducted a group of ablation experiments to validate that our proposed metrics are effective. \\
\textbf{FID.} Fréchet Inception Distance (FID) \cite{heusel2017gans} is often used to evaluate the fidelity of the generated samples in the image generation domain. Iashin and Rahtu \cite{iashin2021taming} also use FID as one metric to evaluate the generated sound. FID is defined by the distance between the distributions of the pre-classification layer’s features of InceptionV3 \cite{szegedy2016rethinking} between fake and real samples, and InceptionV3 is usually pre-trained on ImageNet \cite{deng2009imagenet}.
{\color{black}{The mathematics definition of FID as follow:
\begin{equation}\label{FID}
  ||\boldsymbol{m}_{r}-\boldsymbol{m}_{f}||_{2}^{2} +Tr(\boldsymbol{C}_{r}+\boldsymbol{C}_{f}-2(\boldsymbol{C}_{}\boldsymbol{C}_{f})^{\frac{1}{2}})
\end{equation}
where $\boldsymbol{m}_{r}$ and $\boldsymbol{m}_{f}$ denote the mean of features extracted from real and generated samples. $\boldsymbol{C}_{r}$ and $\boldsymbol{C}_{f}$ are the covariance matrix of features extracted from real and generated samples. $Tr$ denotes the trace of the matrix.}}\\
Considering the difference between images and spectrograms, we adapt the InceptionV3 architecture \cite{szegedy2016rethinking} for the spectrogram and train the model on the AudioSet dataset \cite{gemmeke2017audio}. Specifically, we modify the input convolutional layer and change the number of channels from 3 to 1. We do not use the max pooling operations, in order to preserve {\color{black}spectrogram} resolution at higher layers. We train the InceptionV3 on the Audioset with a batch size of 64 mel-spectrograms using Adam optimizer, the learning rate is $3 \times 10^{-3}$ with weight decay is $1 \times 10^{-3}$.\\
\textbf{KL.} For the text-to-sound generation task, one important thing is to evaluate the relevance between generated samples and text descriptions. Considering a sound comprises of multiple events, we can use a pre-trained classifier (pre-trained InceptionV3 on the AudioSet dataset) to get the probability of generated and real samples, and then calculate Kullback-Leibler (KL) divergence between the two probability distributions.  \\
\textbf{Audio caption loss.} Text-to-sound generation task can be seen as a reverse audio caption \cite{wu2019audio,drossos2020clotho} task. Intuitively, if the generated sample has high fidelity and relevance with text description $\boldsymbol{d}$, {\color{black}{the generated sample can be translated to $\boldsymbol{\hat{d}}$ using an audio caption system}}, and the difference between $\boldsymbol{d}$ and $\boldsymbol{\hat{d}}$ should be small. Thus we propose to turn to the metrics in audio caption field for the text-to-sound generation task. Specifically, we first train an audio caption transformer (ACT) \cite{mei2021audio} on AudioCaps dataset \cite{kim2019audiocaps}. We follow the basic model structure proposed in \cite{mei2021audio}. The difference is that we use a $860 \times 80$ log-mel-spectrogram as input. Then we use SPICE \cite{anderson2016spice} and CIDEr \cite{vedantam2015cider}, which are common evaluation metrics for audio caption tasks, to evaluate the quality of generated samples. The SPICE score guarantees the generated captions are semantically {\color{black}close} to the audio, while the CIDEr score guarantees captions are syntactically fluent. The higher SPICE and CIDEr scores indicate better generation quality.
\subsection{The Effectiveness of FID and KL}
We try to validate whether the FID and KL scores can measure the gap between the generated and real samples. To realize this, we randomly choose {\color{black}a set of} audio $\boldsymbol{X}_o$ from the AudioCaps \cite{kim2019audiocaps} validation set and try to generate a new set $\boldsymbol{X}_f$ from three different aspects, respectively: add Gaussian noise, mask part of the audio content, and add interfering sound. Finally, we calculate the FID and KL scores between $\boldsymbol{X}_o$ and $\boldsymbol{X}_f$ to verify whether the FID and KL metrics are sensitive to these factors. The visualization results are shown in Figure \ref{fig:2}. \\
\textbf{Add Gaussian noise.} Intuitively, if the generated sample contains too much noise, which may not be very well. Thus, we add Gaussian noise to $\boldsymbol{X}_o$ to generate a new set $\boldsymbol{X}_f$, and then we calculate the FID and KL scores between $\boldsymbol{X}_o$ and $\boldsymbol{X}_f$. As Figure \ref{fig:2} (a) shows, the FID and KL scores gradually increase when we add more noise, which indicates that FID and KL metrics are sensitive to extra noise. \\
\textbf{Mask part of audio content.} If the generated sample only contains part of the acoustic information compared to the real sample, the generated sample is suboptimal. Thus, we mask part of the audio content to generate a new set $\boldsymbol{X}_f$. As Figure \ref{fig:2} (b) shows, when we gradually increase the proportion of masked parts, FID and KL scores also increase. \\
% and then calculate the FID and KL scores between $\boldsymbol{X}_o$ and $\boldsymbol{X}_f$
%For example, if the text input is ``open the door'', our generated sample should not include `dog bark' or `music' events.
\textbf{Add interfering sound.} The generated sound should not contain irrelevant acoustic information with the text description. We randomly choose an interfering audio from the AudioCaps training set, and then we directly mix the interfering sound with $\boldsymbol{X}_o$ to build a new set $\boldsymbol{X}_f$. According to Figure \ref{fig:2} (c), we can see that the FID and KL scores gradually increase when the interfering sound increases. \\
\begin{figure}[t] \label{fig:2}
  \centering
  \includegraphics[width=\linewidth]{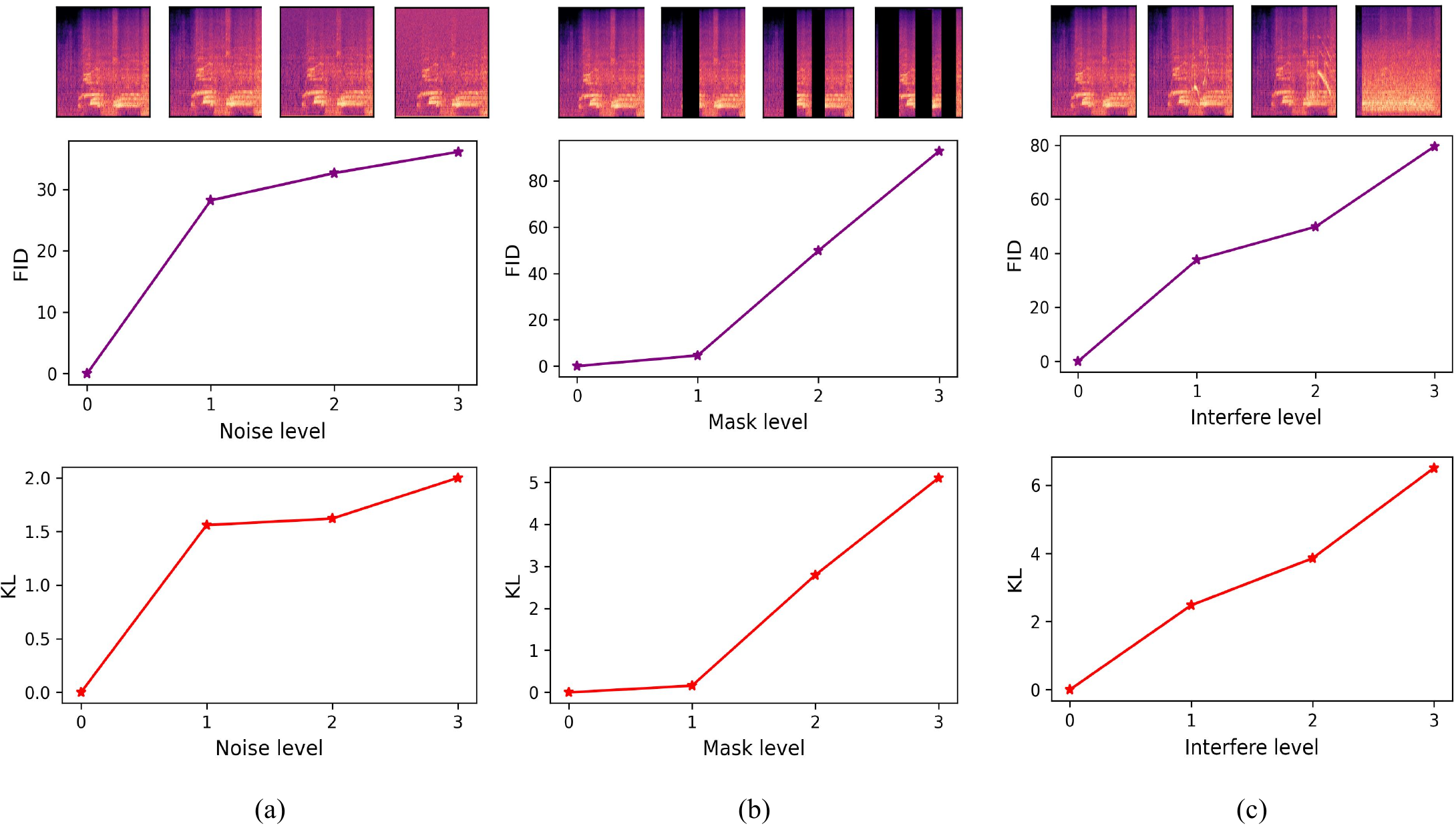}
  \caption{FID and KL are evaluated for \textbf{(a)}: add Gaussian noise, \textbf{(b):} mask part of audio content, \textbf{(c):} mix with other interfere sound. The first row is the simple visualization of how the spectrograms changed when different disturbance level is used. The disturbance level rises from zero and increases to the highest level. We can see that {\color{black}the} FID and KL scores capture the disturbance level very well.}
  \label{fig:2}
  \vspace*{-\baselineskip}
\end{figure}
\section{Experimental setup} \label{experimental setup}
\begin{table}[t] \centering
\caption{The Mean Opinion Score comparasion between AR and Diffsound models. GT denotes the ground truth sound.}
\label{tab:my-table1}
\begin{tabular}{ccccc}
\hline
Model     & Relevance$\uparrow$ & Fidelity$\uparrow$ & Intelligibility$\uparrow$ & MOS$\uparrow$   \\ \hline
GT        & 4.307     & 4.167    & 3.873           & 4.116 \\ \hline
AR        & 2.747     & 2.7      & 2.913           & 2.786 \\
Diffsound & \textbf{3.833}     & \textbf{3.487}    & \textbf{3.36}            & \textbf{3.56}  \\ \hline
\end{tabular}
\end{table}
\subsection{Implementation Details}
%We find that the codebook size $K$ is a very important parameter, so we set $K \in \{256,512,2048\}$.
{\color{black}Our proposed text-to-sound generation framework is not trained end-to-end.} We train each part separately. For text encoder, we directly use the pre-trained BERT or CLIP models. We first train VQ-VAE and vocoder. Then we train the token-decoder with the help of the text encoder and pre-trained VQ-VAE. Note that the VQ-VAE and text encoder are fixed when we train the token-decoder. In the following, we will introduce the details of network structure and training strategy.
% \subsubsection{Text encoder} According to the discussion of Section \ref{text encoder}, we directly use the pre-trained BERT or CLIP models to extract a vector from the text.
\subsubsection{VQ-VAE} In this study, our VQ-VAE's spectrogram encoder $E_{vq}$, spectrogram decoder $G$, and discriminator $D$ follow the setting of VQ-GAN \cite{esser2021taming,iashin2021taming}, which is a variant version of VQ-VAE \cite{van2017neural}. For codebook $\boldsymbol{Z}$, the dimension of each vector $n_z$ is set as 256, and the codebook size $K$ is set as 256. VQ-VAE converts $860 \times 80$ spectrogram into $53 \times 5$ tokens. We train our VQ-VAE model on AudioCaps and AudioSet datasets. We find that training on AudioSet can achieve better performance. Unless specifically stated, we default to using the VQ-VAE pre-trained on AudioSet. The learning rate is fixed and determined as a product of a base learning rate, a number of GPUs, and a batch size. In our experiments, the base learning rate is set as $1 \times 10^{-6}$, and Adam optimizer \cite{kingma2014adam} is used. We train VQ-VAE with batches of 20 mel-spectrograms on 8 Nvidia V100 GPUs. The training takes about 9 days on the AudioSet. To stabilize the training procedure, we zero out the adversarial part of the loss in formula (\ref{SpecVQVAE loss}) (set $\lambda_d=0$) for the first 2 training epochs, after that the adversarial loss is used, and $\lambda_d=0.8$. 
\subsubsection{Autoregressive token-decoder} Inspired by the success in autoregressive sound generation \cite{iashin2021taming,liu2021conditional}, we follow the SOTA backbone of the video-to-sound generation task, and employ a transformer-based network to learn the mapping from text to the spectrogram tokens. Specifically, the autoregressive token-decoder is a 19-layer 16-head transformer with a dimension of 1024. We use one dense layer to map the text features into the transformer’s hidden dimension space (1024), so that the text features can be forward into the transformer. The output of the transformer is passed through a $K$-way softmax classifier. The base learning rate is $1 \times 10^{-6}$, and the AdamW optimizer \cite{loshchilov2017decoupled} is used. The batch size is set as 16 for each GPU. The model is trained until the loss on the validation set has not improved for 2 consecutive epochs. Training the autoregressive token-decoder takes about 2 days on 8 Nvidia P40 GPUs. 
\subsubsection{Diffsound model} For a fair comparison with the autoregressive token-decoder under similar parameters, we also built a 19-layer 16-head transformer with a dimension of 1024 for the Diffsound model. Each transformer block contains a full attention, a cross attention to combine text features and a feed-forward network block. The current timestep $t$ is injected into the network with Adaptive Layer Normalization \cite{ba2016layer}(AdaLN) operator. \\
\textbf{Noise schedule setting.} For the uniform transition matrix, we linearly increase $\overline{\beta}_t$ from 0 to 0.1, and decrease $\overline{\alpha}_t$ from 1 to 0. For the mask transition matrix, we linearly increase $\overline{\gamma}_t$ from 0 to 1, and decrease $\overline{\beta}_t$ from 1 to 0. For the mask and uniform transition matrix, we linearly increase $\overline{\gamma}_t$ and $\overline{\beta}_t$ from 0 to 0.9 and 0.1, and decrease $\overline{\alpha}_t$ from 1 to 0.\\
\textbf{Training details.} For the default setting, we set timesteps $T = 100$ and loss weight $\lambda = 1e-4$ in formula (\ref{diffu final loss}). The mask and uniform transition matrix is used, because we find it can get the best generation performance. We optimize our network using AdamW \cite{loshchilov2017decoupled} with $\beta_1 = 0.9$ and $\beta_2 = 0.94$. The basic learning rate is $3 \times 10^{-6}$, and batch size is 16 for each GPU. We train the Diffsound model on the AudioCaps, which takes about 2 days on 16 Nvidia V100 GPUs (the number of training epochs is set to 400). If we pre-train the Diffsound model on the AudioSet, we use 32 Nvidia V100 GPUs, which takes about 8 days (the number of training epochs is set to 200).
\begin{table}[t] \centering
\caption{The objective metrics comparison between AR decoder and Diffsound. CB denotes that we train the codebook on AudioSet or AudioCaps (Caps for short) datasets. TE denotes the type of text encoder.}
\label{tab:my-table2}
\begin{tabular}{ccccccc}
\hline
Model                      & CB &TE  & FID$\downarrow$      & KL$\downarrow$      & SPICE$\uparrow$  & CIDEr$\uparrow$  \\ \hline
\multirow{2}{*}{AR}        & Caps &BERT & 18.01     & 6.8   & 0.055      & 0.1  \\
                           & Caps &CLIP & 17.94     & 5.98   & 0.082      & 0.2  \\
                           & AudioSet &CLIP & 16.87    & 5.31   & 0.088      & 0.22  \\ \hline
\multirow{2}{*}{Diffsound} & Caps &CLIP & 13.47      & 4.95   & 0.093     & 0.28 \\
                           & AudioSet &CLIP  & \textbf{9.76} & \textbf{4.21}  & \textbf{0.103} & \textbf{0.36}   \\ \hline
\end{tabular}
\end{table}
\begin{table*}[t] \centering
\caption{The generation speed comparison between the AR decoder and Diffsound. Timesteps $T$ denotes the start step in inference stage. Time stride $\Delta_t > 1$ indicates that we use fast inference strategy.}
\label{tab:my-table3}
\begin{tabular}{cccccccc}
\hline
Token-decoder     & Timestep ($T$) & Time stride $\Delta_t$ & FID ($\downarrow$)& KL ($\downarrow$)     & SPICE ($\uparrow$)  & CIDEr ($\uparrow$)  & Speed(spec/s) ($\downarrow$) \\ \hline
AR                                               & -                    & -          & 16.87 & 5.31 & 0.054 & 0.16  & 23.24         \\ \hline
\multicolumn{1}{c|}{\multirow{12}{*}{Diffsound}} & \multirow{4}{*}{25}  & 7         & 16.76 & 4.68 & 0.088 & 0.25  & \textbf{0.53}          \\
\multicolumn{1}{c|}{}                            &                      & 5         & 16.51 & 4.7  & 0.088 & 0.26  & 0.59          \\
\multicolumn{1}{c|}{}                            &                      & 3         & 12.54 & 4.38 & 0.099 & 0.32  & 0.72          \\
\multicolumn{1}{c|}{}                            &                      & 1         & 10.91 & 4.2  & 0.104 & 0.34  & 1.49          \\ \cline{2-8} 
\multicolumn{1}{c|}{}                            & \multirow{4}{*}{50}  & 7         & 15.26 & 4.66 & 0.094 & 0.29  & 0.67          \\
\multicolumn{1}{c|}{}                            &                      & 5         & 14.04 & 4.54 & 0.092 & 0.28  & 0.82          \\
\multicolumn{1}{c|}{}                            &                      & 3         & 11.06 & 4.25 & 0.102 & 0.32  & 1.15          \\
\multicolumn{1}{c|}{}                            &                      & 1         & 10.48 & 4.24 & 0.104 & 0.35  & 2.77          \\ \cline{2-8} 
\multicolumn{1}{c|}{}                            & \multirow{4}{*}{100} & 7         & 11.87 & 4.35 & 0.103 & 0.34  & 1.02          \\
\multicolumn{1}{c|}{}                            &                      & 5         & 12.71 & 4.44 & 0.095 & 0.30  & 1.28          \\
\multicolumn{1}{c|}{}                            &                      & 3         & 10.13 & 4.3  & 0.103 & 0.35  & 1.86          \\
\multicolumn{1}{c|}{}                            &                      & 1         & \textbf{9.76}  & \textbf{4.21} & \textbf{0.103} & \textbf{0.36}  & 4.96          \\ \hline
\end{tabular}
\end{table*}
\subsubsection{Vocoder.} We rely on the official implementation of the MelGAN \cite{kumar2019melgan}. During training, the model
inputs a random sequence of 8192 audio samples (the sample rate is 22050). The vocoder is trained for 200 epochs with a batch size of 256 mel-spectrograms on one P40 GPU for approximately 20 days. {\color{black}Considering the time complexity, we do not use all of the AudioSet data, we randomly choose 40\% audio clips to train the MelGAN.} \\
\subsubsection{The duration of the generated sound}
Consider that each of the clips from the AudioCaps dataset contains 10 seconds of audio. We fixed the duration of the generated sound to 10 seconds to ensure a fair comparison of generated and real sound. {\color{black}As a result, the number of generated mel-spectrogram tokens is fixed at 265 for both the AR token-decoder and the Diffsound model (10 seconds audio corresponding to $80 \times 860$ mel-spectrogram, and the spectrogram encoder in the VQ-VAE model including 16 downsampling operation for both time and frequency dimensions, thus 10 seconds audio can be approximated to $5 \times 53 = 265$ tokens).}
\begin{table}[t] \centering
\caption{Ablation study for three different transition matrices. Furthermore, we also discuss the effect of the final mask rate $\overline{\gamma}_T$ on mask and uniform matrix. U denotes the uniform matrix, M denotes the mask matrix.}
\label{tab:my-table4}
\begin{tabular}{ccccccc}
\hline
ID & Matrix                   & Mask rate & FID$\downarrow$   & KL$\downarrow$   & SPICE$\uparrow$ & CIDEr$\uparrow$ \\ \hline
1  & U                        & 0         & 10.14 & 4.31 & 0.101 & 0.35  \\ \hline
2  & \multirow{5}{*}{MU} & 0.1       & 10.63 & 4.47 & 0.093 & 0.29  \\
3  &                          & 0.3       & 10.64 & 4.35 & 0.099 & 0.34  \\
4  &                          & 0.5       & 10.75 & 4.31 & 0.096 & 0.32  \\
5  &                          & 0.7       & 9.84  & 4.37 & 0.102 & 0.34  \\
6  &                          & 0.9       & \textbf{9.76}  & \textbf{4.21} & \textbf{0.103} & \textbf{0.36}  \\ \hline
7  & M                        & 1         & 11.5  & 4.46 & 0.103 & 0.34  \\ \hline
\end{tabular}
\end{table}
% \begin{table}[t] \centering
% \caption{Ablation study on the effect of the final mask rate $\overline{\gamma}_T$ on mask and uniform matrix. Where $\overline{\gamma}_T=0$ denotes that we use Uniform matrix, $\overline{\gamma}_T=1$ denotes that we use Mask matrix.}
% \label{tab:my-table7}
% \begin{tabular}{cccccc}
% \hline
% Model     & Mask rate $\overline{\gamma}_T$ & FID$\downarrow$   & KL$\downarrow$   & SPICE$\uparrow$ & CIDEr$\uparrow$ \\ \hline
% \multicolumn{1}{c|}{\multirow{7}{*}{Diffsound}} & 0         & 10.14 & 4.31 & 0.101 & 0.35  \\
% \multicolumn{1}{c|}{}                           & 0.1       & 10.63 & 4.47 & 0.093 & 0.29  \\
% \multicolumn{1}{c|}{}                           & 0.3       & 10.64 & 4.35 & 0.099 & 0.34  \\
% \multicolumn{1}{c|}{}                           & 0.5       & 10.75 & 4.31 & 0.096 & 0.32  \\
% \multicolumn{1}{c|}{}                           & 0.7       & 9.84  & 4.37 & 0.102 & 0.34  \\
% \multicolumn{1}{c|}{}                           & 0.9       & \textbf{9.76}  & \textbf{4.21} & \textbf{0.103} & \textbf{0.36}  \\
% \multicolumn{1}{c|}{}                           & 1         & 11.5  & 4.46 & 0.103 & 0.34  \\ \hline
% \end{tabular}
% \end{table}
% \begin{figure}[t]
%   \centering
%   \includegraphics[width=0.9\linewidth]{compare.pdf}
%   \caption{The visualization of reconstructed mel-spectrograms by different codebook size $K$ of VQVAE. The VQVAE model is trained on Audioset dataset.}
%   \label{fig:4}
%   \vspace{-\baselineskip}
% \end{figure}
\section{RESULTS AND ANALYSIS}
In this section, we conduct experiments to verify the effectiveness of our text-to-sound generation framework. Table \ref{tab:my-table1} shows the MOS comparison between the generated and real sound. {\color{black}We can see that our text-to-sound generation framework could achieve good MOS performance (the MOS both large than 2.5) regardless of whether the token-decoder is AR or Diffsound model.} Let's start with a detailed comparison between the AR token-decoder and our proposed Diffsound model. Then we conduct ablation studies for the Diffsound model. 
% Moreover, we explore the effectiveness of pre-training strategy for the Diffsound. Lastly, we use audio caption loss to choose high-relevance sound with the text description.
% \subsection{MOS comparison between the AR and Diffsound models}
% Table \ref{tab:my-table0} shows the MOS comparison between the AR and Diffsound models. We can see that our Diffsound model significantly improve the MOS compared to AR model, \textit{e.g.} MOS 3.56 \textit{v.s} 2.786. By comparing the MOS of Ground Truth and Diffsound model, we see that our method get a good results in relevance and intelligibility aspects. Due to many audio clips includes background noise, we can see that the intelligibility of Ground Truth is low. In the following, we will evaluate the AR and Diffsound models using objective metrics, \textit{e.g.} FID, KL, and audio caption loss.
% \subsection{The performance analysis of VQ-VAE}
% VQ-VAE aims to compress the spectrogram into a group of tokens, and then recover the spectrogram from the tokens. The performance of VQ-VAE significantly influence the generation process. To evaluate the performance of VQ-VAE, we visualize the input mel-spectrogram and the reconstructed mel-spectrogram by VQ-VAE. As Figure \ref{fig:3} shows, we can see that the reconstructed spectrograms by the VQ-VAE model trained on Audioset are better than that trained on the Audiocaps. 
\subsection{The comparison between the AR decoder and Diffsound}
\subsubsection{Subjective and objective metrics}
Table \ref{tab:my-table1} shows the subjective metrics (MOS) comparison between the AR token-decoder and Diffsound model. We can see that our Diffsound model significantly improves the MOS compared to the AR token-decoder, \textit{e.g.}, MOS 3.56 \textit{v.s} 2.786. Due to many audio clips including background noise, the intelligibility of Ground Truth is relatively low.
Table \ref{tab:my-table2} shows the objective metrics comparison between the AR decoder and the Diffsound. Firstly, by comparing rows 1 and 2, we can see that using the CLIP model as the text encoder brings better generation performance than the BERT model. Secondly, we can see that using the VQ-VAE trained on the AudioSet dataset brings better generation quality than on the AudioCaps dataset, we speculate that training on a large-scale dataset can improve the ability of VQ-VAE.
Lastly, by comparing the AR token-decoder and Diffsound model, we can see that the Diffsound model gets better performance on all of the metrics, \textit{e.g.}, FID 9.76  \textit{v.s} 16.87, KL 4.21 \textit{v.s} 5.31, CIDEr 0.36 \textit{v.s} 0.22. In summary, the objective and subjective metrics both indicate the effectiveness of our Diffsound model.
\subsubsection{The generation speed}
Generation speed is also an important metric to evaluate the generation model. To investigate the generation speed between AR token-decoder and Diffsound model, we conducted a group of ablation experiments, and the results are shown in Table \ref{tab:my-table3}. Note that we conducted these experiments on a single Nvidia P40 GPU. We fix the generated sound duration as 10 seconds. {\color{black}We only calculate the time to generate the mel-spectrograms and ignore the vocoder's costs due to the vocoder is the same for all models.} Firstly, we can see that using the AR token-decoder to generate a mel-spectrogram needs about 23 seconds, but using our Diffsound model only takes about 5 seconds. Moreover, our Diffsound model also has better generation performance. Secondly, the generation speed of our Diffsound model can be further improved by using less number of timesteps $T$ and larger time stride $\Delta_t$ but decreases the generation quality. The fastest generation speed is obtained when the number of timesteps $T=25$ and $\Delta_t=7$. The Diffsound model only needs 0.53 seconds to generate a mel-spectrogram, which is 43 times faster than the AR token-decoder with a similar FID score. 
\subsubsection{Visualization}
\begin{figure}[t]
  \centering
  \includegraphics[width=\linewidth]{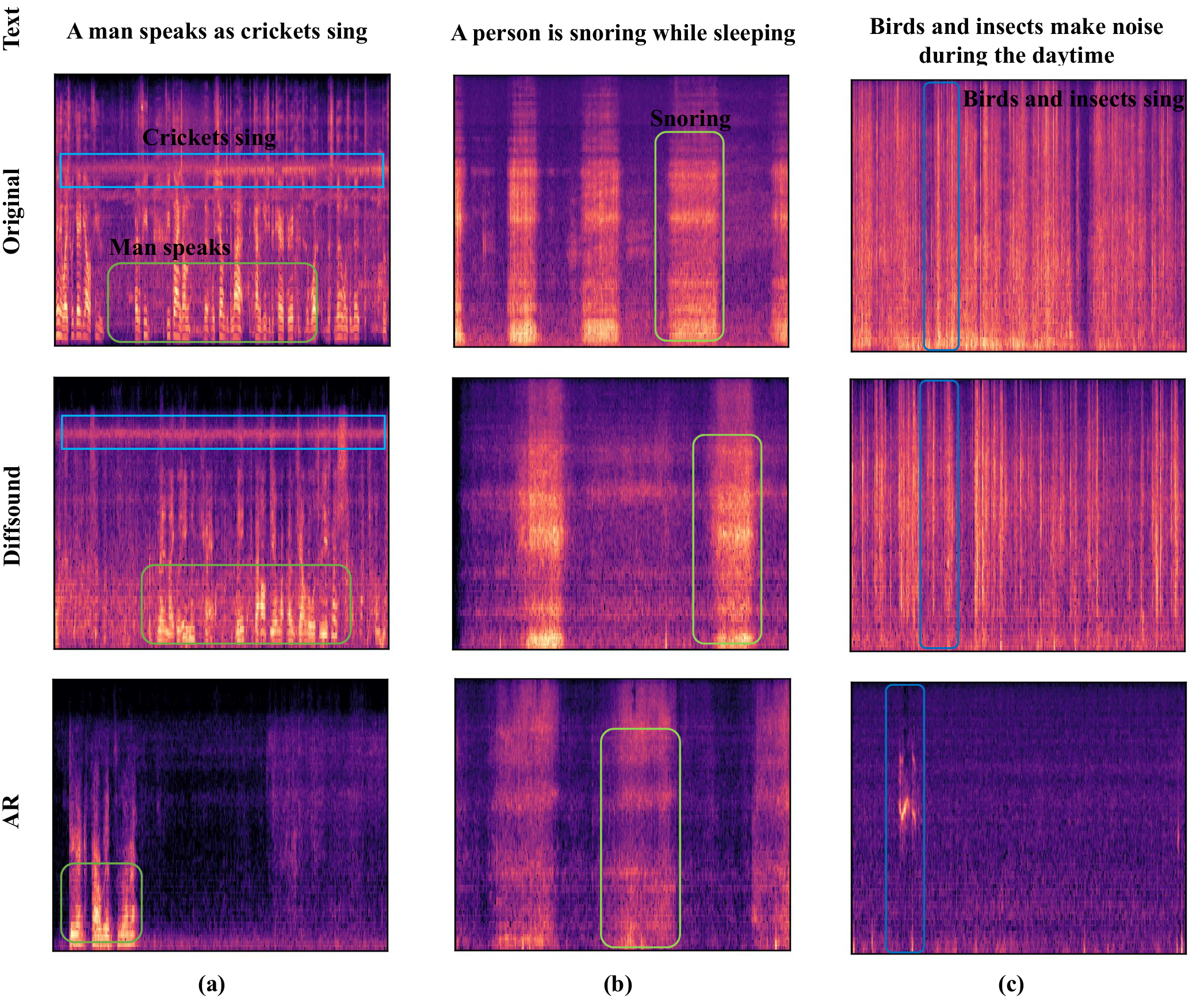}
  \caption{The visualization of generated samples by the Diffsound model and AR token-decoder. The first line is the text input. The second line is the mel-spectrograms of real audio. The last two lines are the mel-spectrograms of generated audio by Diffsound model and AR token-decoder.}
  \label{fig:6}
  \vspace*{-\baselineskip}
\end{figure}

Figure \ref{fig:6} shows some generated samples by the Diffsound model and AR token-decoder. We can see that the Diffsound model can generate a scene with complete semantics compared to the AR token-decoder, \textit{e.g.}, Figure \ref{fig:6} (a) shows that the sound generated by AR token-decoder only includes the man speaks event and the crickets sing is missing. Furthermore, as Figure \ref{fig:6} (b) and (c) show, our Diffsound model has better detailed modelling ability than the AR token-decoder.
\begin{table}[t] \centering
\caption{Ablation study on training and inference steps. Each column uses the same training steps while each row uses the same inference steps. We only report the FID in this table.}
\label{tab:my-table5}
\begin{tabular}{c|cccc}
\hline
                                                                           & \multicolumn{4}{c}{Training step}                                                          \\ \hline
\multirow{4}{*}{\begin{tabular}[c]{@{}c@{}}Inference \\ step\end{tabular}} & \multicolumn{1}{c|}{}    & \multicolumn{1}{c|}{25}    & \multicolumn{1}{c|}{50}    & 100   \\ \cline{2-5} 
                                                                           & \multicolumn{1}{c|}{25}  & \multicolumn{1}{c|}{12.22} & \multicolumn{1}{c|}{11.58}      & 10.91  \\ \cline{2-5} 
                                                                           & \multicolumn{1}{c|}{50}  & \multicolumn{1}{c|}{-}     & \multicolumn{1}{c|}{11.08} & 10.48 \\ \cline{2-5} 
                                                                           & \multicolumn{1}{c|}{100} & \multicolumn{1}{c|}{-}     & \multicolumn{1}{c|}{-}     & \textbf{9.76}  \\ \hline
\end{tabular}
\end{table}
\subsection{Ablation study for Diffsound model}
\subsubsection{Impact of different transition matrices for Diffsound model}
In this section, we explore the impact of three transition matrices on Diffsound model: uniform transition matrix, mask transition matrix, mask and uniform transition matrix. Table \ref{tab:my-table2} presents the results of Diffsound model with three different transition matrices. We can see that the best results are obtained when the mask and uniform matrix is used and $\overline{\gamma}_T=0.9$. 
% \sout{The experimental results verify the effectiveness of the mask and uniform transition matrix: We speculate that only using a uniform transition matrix may bring the problem of semantic change, which leads to the reverse process being hard to learn; And only using the mask transition matrix may make the model tend to focus on the mask token and ignore the context information.} 
{\color{black}From Table~\ref{tab:my-table4}, it can be observed that using the combined mask and uniform transition matrix outperforms using the uniform transition matrix. One possible explanation is that: using a uniform transition matrix may make the reverse process hard to learn; and only using the mask transition matrix may make the model tend to focus on the mask token and ignore the context information.}
\subsubsection{Impact of the final mask rate $\overline{\gamma}_T$ for the mask and uniform transition matrix} We conduct ablation studies to investigate the impact of the final mask rate ($\overline{\gamma}_T$) on the mask and uniform transition matrix. {\color{black}Results are shown on Table \ref{tab:my-table4} (lines 2–6). Experiments show that using $\overline{\gamma}_T=0.9$ outperforms other settings}. 
% If $\overline{\gamma}_T$ is too large (more than 0.9), the Diffsound may pay more attention to the mask token and ignore the context information, so that the performance decreases. Instead, if $\overline{\gamma}_T$ is too small (less than 0.1), the Diffsound also cannot perform well.
\subsubsection{Number of timesteps} We conduct ablation studies to investigate the impact of the number of timesteps $T$ of the training and inference stages for the Diffsound model, with the results shown in Table \ref{tab:my-table5}. In this study, considering the generation speed, we set the maximum number of timesteps $T$ as 100 in both training and inference. We can see that using more number of timesteps in the training and inference stage could get better performance. However using more number of timesteps will cost more time in the inference stage. Furthermore, we can find that it still maintains a good performance when dropping 75 inference steps (\textit{e.g.}, training step is 100, but inference step is 25), which gives us a direction to boost the generation speed. \\
% \usepackage{multirow}
\begin{table}[t] \centering
\caption{The effectiveness of pre-training the Diffsound on Audioset. CL denotes the curriculum learning strategy. PR denotes the pre-training strategy.}
\label{tab:my-table6}
\begin{tabular}{ccccccc}
\hline
Token-decoder     & PR & CL & FID$\downarrow$ & KL$\downarrow$ & SPICE$\uparrow$ & CIDEr$\uparrow$ \\ \hline
\multicolumn{1}{c|}{\multirow{3}{*}{Diffsound}} & \usym{2613}        & \usym{2613}  & 9.76 & 4.21 & 0.103 & 0.36  \\
\multicolumn{1}{c|}{}               & \checkmark        & \usym{2613}  & 8.78 & 4.15 & 0.101 & 0.35  \\
\multicolumn{1}{c|}{}           & \checkmark       & \checkmark  & \textbf{8.27}    & \textbf{4.11}   & \textbf{0.105}      &  \textbf{0.36}  \\ \hline
\end{tabular}
\end{table}
\subsection{The effectiveness of pre-training the Diffsound model on AudioSet}
In this section, we validate whether pre-training the Diffsound on the AudioSet dataset can improve the generation performance. Note that considering the time complexity, we only use about 45\% AudioSet training set. {\color{black}Table \ref{tab:my-table6} shows the experimental results. We can see that using the pre-trained Diffsound model can improve the generation performance in terms of FID and KL (such as lowering the score of FID and KL from 9.76 and 4.21 to 8.78 and 4.15).} We can see that when pre-training and curriculum learning strategies are both used, the best performance is obtained (such as FID score of 8.27, KL score of 4.11, and SPICE score of 0.105), which shows that the curriculum learning strategy is also important. We believe that performance can be further improved when we use more data.
% \usepackage{multirow}
\subsection{Choose high-relevance sound with text description based on audio caption loss}
% According to Section \ref{subsec:evaluation metrix}, we propose to train an audio caption transformer (ACT) on the Audiocaps dataset, and use the ACT to generate text description $\hat{s}$ for the generated sound. SPICE and CIDEr metrics are used to evaluate the gap between $\hat{s}$ and the ground truth $s$. To quickly choose high-relevance samples with the input text, we can rank the samples according to the sum of their SPICE and CIDEr scores, and then we only keep part of them.
Due to the random sample process in the inference stage of the Diffsound model and AR token-decoder, the generated sound may be different in multiple sampling processes even using the same text description. We generated 10 samples for each text in this study. To quickly choose high-relevance samples with the input text, we can rank the samples according to the sum of their SPICE and CIDEr scores, and then we only keep a subset of them. As Table \ref{tab:my-table7} shows, if we only keep the top 2 samples for each text, the AR method's KL score will improve from 5.31 to 5.03, and the Diffsound method's KL score will improve from 4.21 to 3.86. We conjecture that this strategy can help us quickly choose high-relevance samples with the text description. Furthermore, we also observe that the FID score of the Diffsound will slightly increase when top\_k from 10 to 2. We think one of the reasons is that the Diffsound model generates some high-fidelity samples but these samples are irrelevant to the text description.
% \usepackage{multirow}
\begin{table}[t] \centering
\caption{Choose high-relevance sound with text description based on audio caption loss. Top\_k denotes that we keep the top k samples according to the SPICE and CIDEr scores. We totally generate 10 samples for each text examples.}
\label{tab:my-table7}
\begin{tabular}{cccccc}
\hline
Token-decoder     & Top\_k & FID$\downarrow$ & KL$\downarrow$ & SPICE$\uparrow$ & CIDEr$\uparrow$ \\ \hline
\multicolumn{1}{c|}{\multirow{3}{*}{AR}}        & 10     & 16.87 & 5.31 & 0.088 & 0.22  \\
\multicolumn{1}{c|}{}                           & 5      & 16.28 & 5.22 & 0.094 & 0.26  \\
\multicolumn{1}{c|}{}                           & 2       & \textbf{15.72}     & \textbf{5.03}    & \textbf{0.145}    & \textbf{0.41}  \\ \hline
\multicolumn{1}{c|}{\multirow{3}{*}{Diffsound}} & 10     & 9.76  & 4.21 & 0.103 & 0.36  \\
\multicolumn{1}{c|}{}                           & 5      & 9.83  & 4.03 & 0.164 & 0.52  \\
\multicolumn{1}{c|}{}                           & 2     & 10.14   & \textbf{3.86}   & \textbf{0.218}      & \textbf{0.71}   \\ \hline
\end{tabular}
\end{table}
\section{Conclusion} 
In this work, we present a framework for text-to-sound generation tasks, and propose a novel non-autoregressive token-decoder (Diffsound) based on discrete diffusion model, which significantly improves the generation performance and speed compared to the AR token-decoder. We also explore a simple pre-training strategy to further improve the performance of Diffsound model. To effectively evaluate the quality of generated samples, we designed three objective evaluation metrics for this task. Both objective and subjective metrics verified the effectiveness of Diffsound model.

This work still has some limitations that need to be
addressed in our future work, \textit{e.g.}, our generation framework is not end-to-end, we separately train the VQ-VAE, the token decoder, and the vocoder, which may not be optimal. In the future, we will explore an end-to-end sound generation framework.
% Figure \ref{fig:5} shows some examples, we can see that the vocoder will lose some details in the reconstruction process, \textit{e.g.} the green box fixed area on the third column. 

% \section*{Acknowledgments}
% This paper was supported by Shenzhen Science and Technology Fundamental Research Programs JSGG20191129105421211 and GXWD20201231165807007-20200814115301001.

% {\appendix[Proof of the Zonklar Equations]
% Use $\backslash${\tt{appendix}} if you have a single appendix:
% Do not use $\backslash${\tt{section}} anymore after $\backslash${\tt{appendix}}, only $\backslash${\tt{section*}}.
% If you have multiple appendixes use $\backslash${\tt{appendices}} then use $\backslash${\tt{section}} to start each appendix.
% You must declare a $\backslash${\tt{section}} before using any $\backslash${\tt{subsection}} or using $\backslash${\tt{label}} ($\backslash${\tt{appendices}} by itself
%  starts a section numbered zero.)}
\appendices
\section{The Proof of formula (\ref{formula:quick cal uniform matrix})} \label{appendix1}
We use mathematical induction to prove formula (\ref{formula:quick cal uniform matrix}). We have following conditional information:
\begin{equation}\label{formula:proof1}
\begin{aligned}
   \beta_t \in [0,1], \alpha_t=1-K\beta_t,
   \overline{\alpha}_t=\prod_{i=1}^{t}\alpha_i, \overline{\beta}_t=(1- \overline{\alpha}_t)/K.
\end{aligned}
\end{equation}
Now we want to prove that $\overline{\boldsymbol{Q}}_t \boldsymbol{c}(x_0) = \overline{\alpha}_t \boldsymbol{c}(x_0) + \overline{\beta}_t$.
Firstly, when $t=1$, we have:
\begin{equation}\label{formula:proof1}
\overline{\boldsymbol{Q}}_1\boldsymbol{c}(x_0)=
\left\{
             \begin{array}{lr}
             \overline{\alpha}_1+\overline{\beta}_1, & x=x_0  \\
             \overline{\beta}_1, & x \neq x_0\\
             \end{array}
\right.
\end{equation}
which is clearly hold. Suppose the formula (\ref{formula:quick cal uniform matrix}) holds at step $t$, then for $t = t + 1$, we have:
$$
\overline{\boldsymbol{Q}}_{t+1} \boldsymbol{c}(x_0) = \boldsymbol{Q}_{t+1}\overline{\boldsymbol{Q}}_t \boldsymbol{c}(x_0).
$$
Now we consider two conditions: \\
(1) when $x=x_0$ in step $t+1$, we have two situations in step $t$, that is, $x=x_0$ and $x \neq x_0$, so that we have:
\begin{equation}\label{formula:proof2}
\begin{aligned}
  \boldsymbol{Q}_{t+1}\boldsymbol{c}(x_0)_{(x)}&=(\overline{\alpha}_t+\overline{\beta}_t)(\alpha_{t+1}+\beta_{t+1})+(K-1)\overline{\beta_t}\beta_{t+1} \\
  &=\overline{\alpha}_{t+1} +\overline{\alpha}_t\beta_{t+1}+\overline{\beta}_t\alpha_{t+1}+K\overline{\beta}_t\beta_{t+1} \\
  &= \overline{\alpha}_{t+1} + \overline{\alpha}_t\beta_{t+1}+\overline{\beta}_t\alpha_{t+1} + (1-\overline{\alpha}_t)\beta_{t+1} \\
  &= \overline{\alpha}_{t+1} + \frac{(1-\overline{\alpha}_t)}{K}\alpha_{t+1} + \frac{(1-\alpha_{t+1})}{K} \\
%   &= \overline{\alpha}_{t+1} + \frac{(1-\overline{\alpha}_{t+1})}{K} \\
  &= \overline{\alpha}_{t+1} + \overline{\beta}_{t+1}.
\end{aligned}
\end{equation}
(2) when $x \neq x_0$ in step $t+1$, we also have two situations in step $t$, that is, $x=x_0$ and $x \neq x_0$,  so that we have:
\begin{equation}\label{formula:proof3}
\begin{aligned}
  \boldsymbol{Q}_{t+1}\boldsymbol{c}(x_0)_{(x)}&=\overline{\beta}_t(\alpha_{t+1}+\beta_{t+1}) +\overline{\beta_t}\beta_{t+1}(K-1) + \overline{\alpha}_t\beta_{t+1} \\
  &=\overline{\beta}_t(\alpha_{t+1}+\beta_{t+1}+\beta_{t+1}(K-1))+
  \overline{\alpha}_t\beta_{t+1} \\
  &= \overline{\beta}_t +  \overline{\alpha}_t\beta_{t+1} \\
  &= \frac{(1-\overline{\alpha}_t)}{K} + \frac{\overline{\alpha}_t(1-\alpha_{t+1})}{K} \\
%   &=  \frac{1-\overline{\alpha}_{t+1}}{K} \\
  &= \overline{\beta}_{t+1}.
\end{aligned}
\end{equation}
The proof of formula (\ref{formula:quick cal uniform matrix}) is completed. \\
Furthermore, we can see that when $t$ is enough large, $\overline{\alpha}_t$ is close to 0. The formula (\ref{formula:quick cal uniform matrix}) changes to $\overline{\boldsymbol{Q}}_t \boldsymbol{c}(x_0) = 1/K$. Thus we can derive the stationary distribution $p(\boldsymbol{x}_T)=[1/K, 1/K, \cdots, 1/K]$.
\section{The Proof of formula (\ref{formula:quick cal mask matrix})} \label{appendix2}
We also use mathematical induction to prove formula (\ref{formula:quick cal mask matrix}). We have following conditional information:
\begin{equation}\label{formula:proof4}
\begin{aligned}
   \gamma_t \in [0,1], \beta_t=1-\gamma_t ,
   \overline{\gamma}_t= 1- \overline{\beta}_t, \overline{\beta}_t=\prod_{i=1}^{t} \beta_i.
\end{aligned}
\end{equation}
Now we want to prove that $\overline{\boldsymbol{Q}}_t \boldsymbol{c}(x_0) = \overline{\beta}_t \boldsymbol{c}(x_0) + \overline{\gamma}_t\boldsymbol{c}(K+1)$.
Firstly, when $t=1$, we have:
\begin{equation}\label{formula:proof5}
\overline{\boldsymbol{Q}}_1\boldsymbol{c}(x_0)=
\left\{
             \begin{array}{lr}
             \overline{\beta}_1, & x=x_0  \\
             \overline{\gamma}_1, & x = K+1\\
             \end{array}
\right.
\end{equation}
which is clearly hold. Suppose the formula (\ref{formula:quick cal mask matrix}) is hold at step $t$, then for $t = t + 1$, we have:
$$
\overline{\boldsymbol{Q}}_{t+1} \boldsymbol{c}(x_0) = \boldsymbol{Q}_{t+1}\overline{\boldsymbol{Q}}_t \boldsymbol{c}(x_0)
$$
Now we consider two conditions: \\
(1) when $x=x_0$ in step $t+1$, we only have one situation in step $t$, that is, $x=x_0$, so that we have:
\begin{equation}\label{formula:proof6}
\begin{aligned}
  \boldsymbol{Q}_{t+1}\boldsymbol{c}(x_0)_{(x)}&=\beta_{t+1}\overline{\beta}_t \\
  &= \overline{\beta}_{t+1}.
\end{aligned}
\end{equation}
(2) when $x = K+1$ in step $t+1$, we have two situations in step $t$, that is, $x=x_0$ and $x = K+1$, so that we have:
\begin{equation}\label{formula:proof7}
\begin{aligned}
  \boldsymbol{Q}_{t+1}\boldsymbol{c}(x_0)_{(x)}&=\overline{\gamma}_t + \overline{\beta}_t\gamma_{t+1} \\
  &= 1-\overline{\beta}_t + \overline{\beta}_t\gamma_{t+1} \\
  &= 1-\overline{\beta}_t(1-\gamma_{t+1}) \\
  &= 1-\overline{\beta}_{t+1} \\
  &= \overline{\gamma}_{t+1}.
\end{aligned}
\end{equation}
The proof of formula (\ref{formula:quick cal mask matrix}) is completed.\\
Similarly, when $t$ is enough large, $\overline{\beta}_t$ is close to 0. The formula (\ref{formula:quick cal mask matrix}) can be written as $\overline{\boldsymbol{Q}}_t \boldsymbol{c}(x_0) =  \overline{\gamma}_t\boldsymbol{c}(K+1)$, where $\overline{\gamma}_t=1$. Thus we can derive the stationary distribution $p(\boldsymbol{x}_T)=[0, 0, \cdots, 1]$. 
\bibliographystyle{ieeetr}
\normalem
\balance
\bibliography{refs.bib}
\end{document}